\documentclass[conference]{IEEEtran}
\pdfoutput=1
\IEEEoverridecommandlockouts
\usepackage{xcolor}
\usepackage{colortbl}
\usepackage{bbding}
\usepackage{enumitem}
\usepackage{cite}
\usepackage[hidelinks]{hyperref}
\usepackage[cmex10]{amsmath}
\usepackage[capitalise,noabbrev]{cleveref}

\usepackage{makecell}

\usepackage{lipsum}

\usepackage{booktabs,tabularx,ragged2e}

\usepackage[pdftex]{graphicx}

\usepackage{amssymb}

\usepackage{array}

\usepackage{mdwmath}
\usepackage{mdwtab}
\usepackage{url}

\usepackage{tikz}
\newcommand*\emptycirc[1][1ex]{%
  \begin{tikzpicture}[baseline=-3]
  \draw (0,0) circle (#1);
  \end{tikzpicture}}
\newcommand*\halfcirc[1][1ex]{%
  \begin{tikzpicture}[baseline=-3]
  \draw[fill] (0,0)-- (90:#1) arc (90:270:#1) -- cycle ;
  \draw (0,0) circle (#1);
  \end{tikzpicture}}
\newcommand*\fullcirc[1][1ex]{%
  \begin{tikzpicture}[baseline=-3]
  \draw[fill] (0,0) circle (#1);
  \end{tikzpicture}}
\newcommand*\crosscirc[1][1ex]{%
\begin{tikzpicture}[baseline=-3]
\draw (0,0) circle (#1);
\draw (-#1,-0.0#1) -- (#1,-0.0#1);
\draw (+0.00#1,-#1) -- (+0.00#1,#1);
\end{tikzpicture}}

\newcommand*\yes[1][1ex]{\fullcirc}
\newcommand*\unclear[1][1ex]{\halfcirc}
\newcommand*\no[1][1ex]{\emptycirc}
\newcommand*\notapplicable[1][1ex]{\crosscirc}

\newcommand*\circled[1]{\tikz[baseline=(char.base)]{
            \node[shape=circle,draw,inner sep=1pt] (char) {\footnotesize\textbf{#1}};}}

\usepackage{mdframed}

\usepackage[nolist]{acronym}

\begin{document}

\title{Mens Sana In Corpore Sano:\\Sound Firmware Corpora for Vulnerability Research}

\author{\IEEEauthorblockN{René Helmke\IEEEauthorrefmark{1},
Elmar Padilla\IEEEauthorrefmark{1}, and 
Nils Aschenbruck$^{\circ}$}
\IEEEauthorblockA{\IEEEauthorrefmark{1}Fraunhofer FKIE, Cyber Analysis \& Defense, Germany, \{firstname.lastname\}@fkie.fraunhofer.de}
\IEEEauthorblockA{$^{\circ}$Osnabrück University, Distributed Systems Group, Germany, aschenbruck@uos.de}}

\maketitle

\begin{abstract}
Firmware corpora for vulnerability research should be \textit{scientifically sound}.
Yet, several practical challenges complicate the creation of sound corpora:
Sample acquisition, e.g., is hard and one must overcome the barrier of proprietary or encrypted data.
As image contents are unknown prior analysis,
it is hard to select \textit{high-quality} samples that can satisfy scientific demands.
Ideally, we help each other out by sharing data.
But here, sharing is problematic due to copyright laws.
Instead, papers must carefully document each step of corpus creation:
If a step is unclear, replicability is jeopardized.
This has cascading effects on result verifiability, representativeness, and, thus, soundness.

Despite all challenges, how can we maintain the soundness of firmware corpora?
This paper thoroughly analyzes the problem space and investigates its impact on research:
We distill practical binary analysis challenges that significantly influence corpus creation.
We use these insights to derive guidelines that help researchers to nurture corpus replicability and representativeness.
We apply them to
44 top tier papers and systematically analyze scientific corpus creation practices.
Our comprehensive analysis confirms that there is currently no common ground in related work.
It shows the added value of our guidelines, as they discover methodical issues in corpus creation and unveil
miniscule step stones in documentation.
These blur visions on representativeness, hinder replicability, and, thus, negatively impact the soundness of otherwise excellent work.

Finally, we show the feasibility of our guidelines and build a new, replicable corpus for large-scale analyses on Linux firmware: LFwC.
We share rich meta data for good (and proven) replicability.
We verify unpacking, deduplicate, identify contents, provide ground truth, and show LFwC's utility for research.
\end{abstract}

\begin{acronym}
\acro{IoT}{Internet-of-Things}
\acro{COTS}{commercial off-the-shelve}
\acro{ISA}{Instruction Set Architecture}
\acro{HIL}{Hardware-In-the-Loop}
\acro{LFwC}{Linux Firmware Corpus}
\acro{FACT}{Firmware Analysis and Comparison Tool}
\acro{OS}{operating system}
\acro{MMU}{memory management unit}
\end{acronym}

\section{Introduction}
Embedded systems are part of everyone's life.
Their proliferation in households, industries, and critical domains makes them highly lucrative targets for cyber attacks with devastating impact, e.g.,~\cite{mirai,stuxnet,vpnfilter,pipedream}.
Thus, finding vulnerabilities in the firmware running on these systems is an important task.

Firmware vulnerability research becomes a matryoshka doll of nested analysis problems when no source code is available:
Acquisition is hard, data may be encrypted, and architectures are manifold~\cite{qasem_vulnerability_detection_survey}.
Heterogeneity and resource constraints defy established analysis methods for general-purpose systems~\cite{challenges_firmware_rehosting}.

Automated firmware vulnerability research has, thus, become a prevalent research topic~\cite{challenges_firmware_rehosting,qasem_vulnerability_detection_survey}:
Common static methods are cross-platform code similarity or taint analysis~\cite{firmup,rediniKaronteDetectingInsecure2020}.
Dynamic approaches explore scalable emulation to create test beds for techniques like fuzzing~\cite{challenges_firmware_rehosting}.

Regardless of the method conducted, there is a need for high-quality firmware corpora for sound evaluations.
This is intuitive, as related fields show that careful curation, rich meta data, and meticulous documentation foster scientific rigor, enable replicability, and emulate real conditions~\cite{plohmann2017malpedia,klees}.

We have conducted a literature review on firmware corpora and found little consensus on their creation:
Some researchers, e.g., scrape the Internet~\cite{costinLargeScaleAnalysisSecurity2014,chenAutomatedDynamicAnalysis2016} while others select few images~\cite{eschweilerDiscovREEfficientCrossArchitecture2016}.
Some describe unpacking~\cite{rediniKaronteDetectingInsecure2020}; others do not~\cite{eschweilerDiscovREEfficientCrossArchitecture2016}.
Some collect product data~\cite{chenAutomatedDynamicAnalysis2016}.
Others do not~\cite{costinLargeScaleAnalysisSecurity2014}.
Copyright and intellectual property laws limit sample sharing.
Thus, some share no data~\cite{muenchWhatYouCorrupt2018}, but others provide source links~\cite{costinLargeScaleAnalysisSecurity2014}.

All of the above affects soundness.
Unpacking is an example~\cite{qasem_vulnerability_detection_survey}:
If we share too few details, replication may fail.
This may push us towards small corpora, which can add bias.
We may bulk collect to improve unpacking odds;
but without filters, we affect representativeness:
If data is riddled with, e.g., old samples, it may no longer represent today's vulnerabilities. 

To sidestep these problems, we could craft synthetic tests, but they can not fully model real conditions~\cite{geng_empirical_2020}.
Thus, this paper focuses on real firmware.
Of course, analysis methods affect our corpora;
not all work targets the same systems.
But ideally, we may agree on unified and sound data requirements.
This paper is a comprehensive analysis of the problem space of sound firmware corpora.
We provide guidelines to improve their soundness, and contribute a new corpus that follows these guidelines.
More specifically, our contributions are as follows:
\begin{itemize}
    \item We distill common firmware analysis problems from related work~\cite{challenges_firmware_rehosting,qasem_vulnerability_detection_survey} to pinpoint corpus creation challenges:
    What makes the creation of scientifically sound corpora so hard?
    We set ground for a practical perspective on (and better understanding of) the problem space.
    \item With the challenges in mind, we propose a framework of data requirements to increase the soundness of firmware corpora:
    Three superordinate goals
    are nurtured by six
    requirements and 16 measures.
    It can support researchers by raising attention to the small step stones that lie in their way towards scientifically sound corpora.
    As a Latin idiom states: \textit{Mens sana in corpore sano} -- 
    These are guidelines to support \textit{a healthy mind in a healthy body}.
    \item We show that there is no common ground on corpus documentation, even in otherwise excellent work:
    We review 44 top tier papers to collect data on our framework.
    We discover that missing meta data, scarce documentation, and inflated corpus sizes blur visions on representativeness and hinder replicability.
    This demonstrates the need for a set of best corpus practices that the community can agree on, highlights the added value of our framework, and defines the research gap that this analysis paper fills.
    \item We describe a new \ac{LFwC} to show our framework's feasibility.
    It contains 10,913 high-quality, meticulously documented, and fully unpackable images from ten manufacturers.
    It covers 2,365 devices and 22 device classes.
    We share rich meta data with the community and perform deduplication and content identification.
    We use an established open source tool for replicable, effective, and verified unpacking.
    The scripts are available at \url{https://github.com/fkie-cad/linux-firmware-corpus}.
    Access to the meta data can be requested on Zenodo~\cite{lfwc_full_zenodo}.
\end{itemize}

\section{An Analysis of Corpus Creation Challenges}
\label{sec:challenges}
Establishing a practice-oriented perspective on the problem space of scientific firmware corpora allows us to propose sane data requirements and conduct fair analyses of existing work.
We consult surveys by Wright et al.~\cite{challenges_firmware_rehosting} and Qasem et al.~\cite{qasem_vulnerability_detection_survey}, who identify binary analysis challenges from two perspectives:
The former targets vulnerability research and the latter emulation.
Yet, explicit questions on \textit{which} and \textit{how} challenges affect corpora remain unanswered.

We approach these questions by collecting and analyzing challenges from the surveys~\cite{qasem_vulnerability_detection_survey,challenges_firmware_rehosting}:
We deduplicate similar items, reorganize them, and distill their impact on corpus creation by including examples from related work.
We drop challenges with no clear impact.
Superordinate goals lack clarity, e.g., \textit{improving accuracy}~\cite{qasem_vulnerability_detection_survey}.
We mark \emph{General} and \emph{Method-Specific} challenges that do not apply to all papers.

We use the taxonomy by Muench et al.~\cite{muenchWhatYouCorrupt2018} for firmware classification.
They define four types on the axes of \ac{OS} abstraction and specialization:
general purpose (Type-0), retrofitted general purpose (I), special purpose (II), and bare-metal~(III).
Respective examples are Windows, a retrofitted Linux, VxWorks, and bare-metal binaries based on Contiki-NG.
Appendix \ref{appendix:types} describes the types in detail.

\begin{figure}[t!]
    \begin{center}
        \includegraphics[width=1.\linewidth]{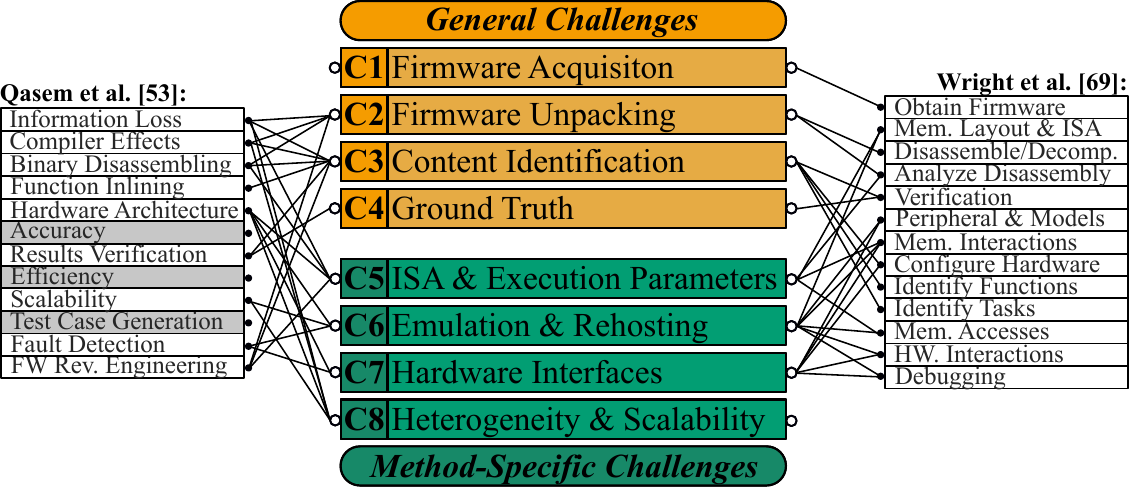}
        \caption{Firmware analysis challenges that can have an impact on scientific firmware corpus construction. We distill common challenges from the surveys of Wright et al.~\cite{challenges_firmware_rehosting} and Qasem et al.~\cite{qasem_vulnerability_detection_survey} and group them into eight descriptive classes.
        Items from the source surveys that show no clear impact on corpus creation are in grey: \emph{Accuracy}, \emph{Test Case Generation}, and \emph{Efficiency}. We mark \textit{General} and \textit{Method-Specific} challenges.}
        \label{fig:analysis_challenges}
    \end{center}
\end{figure}

\cref{fig:analysis_challenges} shows the distilled challenges with impact on corpus creation.
The first four are general, the latter four are method-specific.
The connections show all source challenges with their newly associated descriptive class.

\begin{figure*}[t!]
    \begin{center}
        \includegraphics[width=0.95\textwidth]{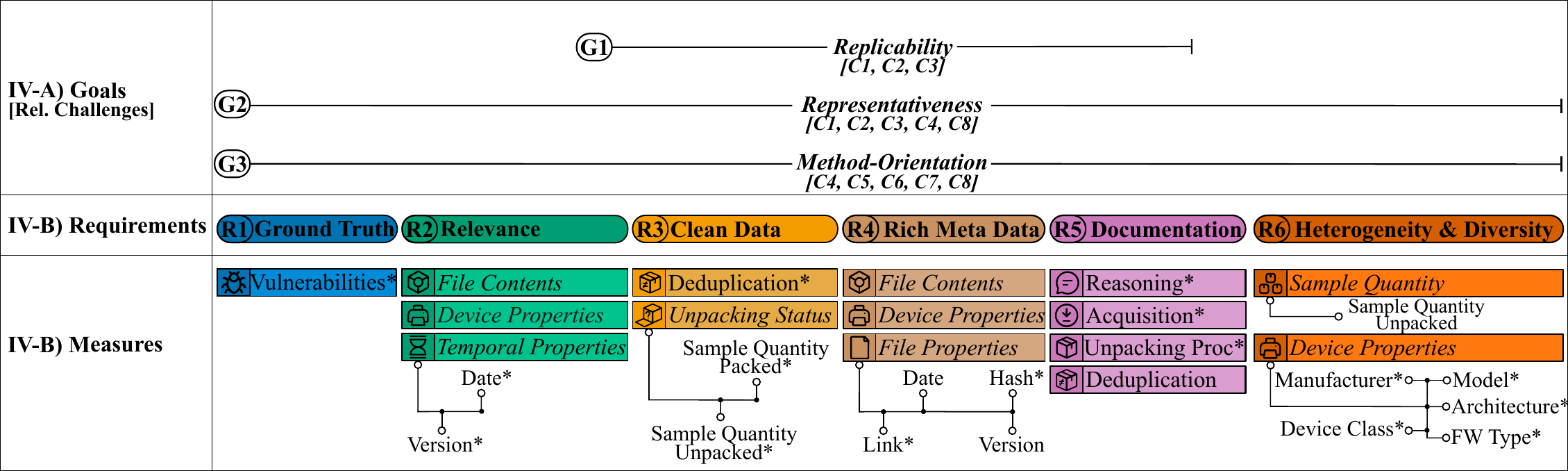}
        \caption{A framework of guidelines for the creation of scientifically sound firmware corpora. It consists of three layers: On top are the superordinate scientific goals \textbf{Replicability}, \textbf{Representativeness}, and \textbf{Method-Orientation}. They are associated with the identified problems in corpus creation from~\cref{sec:challenges}. To achieve these goals, a corpus must fulfill six requirements, which is the second layer: \textbf{Ground Truth}, \textbf{Relevance}, \textbf{Clean Data}, \textbf{Rich Meta Data}, \textbf{Documentation}, and \textbf{Heterogeneity \& Diversity}. In layer three, we add a list of 16 unique and practical measures designated by an asterisk (*). They help to assess the fulfillment of the previously mentioned requirements. Note that measures can serve multiple requirements. Abstract measures are written in \textit{cursive}. We do not claim list completeness, as varying paper scenarios and method constraints may imply additional or substitute measures.}
        \label{fig:requirements}
    \end{center}
\end{figure*}

\begin{itemize}[align=left, wide=0em, leftmargin=1em]
    \item[\textbf{C1}] \textit{Firmware Acquisition.} Empirical data shows high Type-I accessibility~\cite{rediniKaronteDetectingInsecure2020,firmup,chenAutomatedDynamicAnalysis2016,costinLargeScaleAnalysisSecurity2014,costin_dynamic,genius,elsabaghFIRMSCOPEAutomaticUncovering2020,angelakopoulosFirmSoloEnablingDynamic2023}:
    Research scrapes the Internet to download samples from manufacturer or third party sites.
    This is non-trivial as scrapers must navigate volatile pages and links.
    If access is restricted, which is common when entering industrial settings with Type-II and -III systems,    
    one resorts to \ac{HIL} testing or invasive extraction.
    The latter means over-the-air captures, reading debug ports or extracting chips.
    The devices a paper targets dictate feasible acquisition and corpus sizes.
    Yet, this is not bound to firmware types.
    E.g., some Type-I firmware remains unmaintained and unpublished: Scalable acquisition is infeasible.
    Vice versa, some manufacturers make Type-III easily accessible.
    \item[\textbf{C2}] \textit{Firmware Unpacking.}
    Researchers must unpack firmware before analyzing its contents.
    But many archive types exist and manufacturers often use proprietary formats or encryption~\cite{costinLargeScaleAnalysisSecurity2014}.
    Thus, there is significant effort to identify, reverse engineer, and decrypt samples.
    The community uses tools that unpack known formats, e.g.,~\cite{binwalk,bat,unblob,fact}.
    Yet, the challenge persists as formats and encryption changes.
    \item[\textbf{C3}] \textit{Content Identification.} Prior to analysis, firmware contents are often unknown.
    This yields two problems:
    First, content serves as unpacking validation.
    If there is no ground truth on contents, there is uncertainty in unpacking success.
    Second, we must ensure that contents match analysis requirements.
    One can not evaluate, e.g., kernel analyses on images without kernel.
    Information loss is a related issue, e.g., binary information is needed but stripped from the image~\cite{qasem_vulnerability_detection_survey}.
    \item[\textbf{C4}] \textit{Ground Truth.} Firmware with ground truth is needed for result validation.
    Samples might have known vulnerabilities of weaknesses targeted by a method, e.g., taint-style bugs for data flow analyses.
    Another example would be firmware with unstripped executables to test correct function identification.
    This a problem, as such data is not always available.
    \item[\textbf{C5}] \textit{\acs{ISA} \& Execution Parameters.} Methods that analyze or execute code must often be aware of the \ac{ISA} and execution parameters like base address, entrypoint, and memory layout~\cite{challenges_firmware_rehosting,qasem_vulnerability_detection_survey}.
    For Type-I and -II, this is a smaller issue as systems use known formats providing these parameters, e.g., ELF.
    Yet, the documentation of proprietary Type-III systems is often restricted and they operate on unknown execution parameters.
    Researchers can only use samples where this information is available.
    \item[\textbf{C6}] \textit{Emulation.} Dynamic methods like fuzzing are faced with emulation challenges~\cite{challenges_firmware_rehosting}:
    Generally, one can only test devices with components supported by the emulators and detail information on the execution environment.
    In all other cases, one must laboriously establish compatibility, or develop strategies to automate the process~\cite{greenhouse,firmae,pretender,chenAutomatedDynamicAnalysis2016}.
    Example implementation tasks are \ac{ISA} translation layers and peripheral behavior, which must consider timing, interrupts, or direct memory access.
    This is especially relevant considering the heterogeneity, specialization, and restricted documentation of Type-II to -III systems.
    \item[\textbf{C7}] \textit{Hardware Interfaces.} \ac{HIL} methods can only work with sample devices where interfacing during runtime is possible.
    This might be an Ethernet port or debug interface such as I2C, JTAG, SPI, or UART on the PCB.
    \item[\textbf{C8}] \textit{Heterogeneity \& Scalability.} \ac{HIL} and its need for real device interaction show that methods are unequally scalable.
    Static analyses, e.g., can entirely work on firmware samples, are not constrained by execution, and, thus, scale better.
    This influences corpus sizes, as we can see when comparing, e.g., 5,000 FirmUp~\cite{firmup} samples with three Avatar~\cite{zaddachAvatarFrameworkSupport2014} samples.
    Heterogeneity and varying method applicability are related:
    Cross-platform code similarity, e.g., scales better across \acp{ISA} than symbolic execution for a few ARM flavors.
\end{itemize}

\section{Requirements for Sound Firmware Corpora}
\label{sec:reqs}

Firmware corpora are the data source we use to assess our binary analysis methods.
Thus, their composition and documentation play important roles in evaluating method performance; they should be \textit{scientifically sound}:
The results we derive from them should be transparent, comprehensible, and verifiable; conclusions one can draw should accurately describe or approach an objective truth and clarify limitations.
With the challenges from~\cref{sec:challenges} in mind, we will now propose a framework of data requirements that serve as practical guidelines for the creation of sound firmware corpora.

\cref{fig:requirements} illustrates a system with three layers:
There are superordinate goals, which are nurtured by six requirements.
The requirements describe properties to consider for corpus creation.
Goals and requirements are rather abstract concepts that may be hard to measure in practice.
For instance, who draws the line between old and actual samples to conclude upon relevance?
Thus, we identified 16 practical \textit{measures} that feed into the system and allow estimations and comparisons on requirement fulfillment.
Intuitively, there is no claim of completeness:
Scenarios may need specialized measures.
Covering all possibilities is infeasible when aiming for generalizability.
Thus, we include items with assumed universal applicability.

\subsection{Goals of Scientifically Sound Firmware Corpora}
\label{sec:reqs:goals}

Three intuitive goals shall steer us towards soundness:
Replicability, Representativeness, and Method-Orientation.
First, a sound corpus provides \textbf{Replicability (G1)}, allowing for independent result verification.
It enables comparability, as the corpus can serve as benchmark for multiple papers.
However, roadblocks hinder accessibility for third parties:
C1 shows that firmware acquisition and sharing is non-trivial due to, e.g., copyright laws.
C2 and C3 show the issues of unpacking, validation, and content identification to parameterize analysis methods and verify ground truth.
With C1 as primary roadblock that complicates sharing of (pre-prepared) samples, the goal of Replicability forces us to apply best effort approaches within our own possibilities:
To meticulously document each aspect of the corpus from acquisition to unpacking, and propagate as much meta data as legally possible.
The latter helps with content identification and improves the odds of successful corpus reconstruction.

Corpora should be \textbf{Representative (G2)}.
Plohmann et al.~\cite{plohmann2017malpedia}, who maintain a scientific malware corpus with similar problem space as firmware corpora, explain representativeness:
It \textit{``means that the selection of samples contained in the data set should be prevalent and suitable for the deduction of results that are of real-world relevance''}\cite{plohmann2017malpedia}.
For firmware corpora, this means that they should adequately model real-world distributions of \textit{relevant} samples with heterogeneous properties like manufacturers and models.
What \textit{relevant} means will be separately discussed in~\cref{sec:reqs:reqs}.
Ideally, we can model market distributions of device properties.
But those are often unknown and C1 complicates sample acquisition, which affects C8.
Another practical, but less accurate, approach is to aim for the largest possible heterogeneity to demonstrate scalability, applicability, impact, and performance of an analysis method.

Malpedia~\cite{plohmann2017malpedia} shares similar problems with firmware corpus creation:
Malware samples are often packed or encrypted, too.
Thus, both share the challenges C2 and C3, affecting ground truth (C4).
This impacts verifiable representativeness; it is hard to check the contents of a box without looking into it.
Thus, we adopt the \textit{quality over quantity} credo~\cite{plohmann2017malpedia}.
By \textit{quality}, we mean that all samples can be unpacked while their contents are known, deduplicated, appropriately match the scenario, and avoid data skew.
Balancing firmware sample quality and quantity is a tightrope act, which becomes apparent when looking at machine learning and its common pitfalls~\cite{arp_ml_cs_best_practices}:
Here, one seeks a sweet spot between quantity for training and quality to avoid overfitting and improve model performance.

Finally, we propose the goal of \textbf{Method-Orientation (G3)}:
Research questions differ, papers target firmware of different availability, and analysis scalability varies.
These aspects ultimately dictate the feasibility of other goals (C5 to C8).
We provide three examples:
First, emulative approaches can only include firmware that is successfully re-hosted, with sufficient fidelity to derive results of real-world relevance.
Second, symbolic execution might be bound to certain \acp{ISA}; there is little use to include MIPS as heterogeneity property when the engine only supports ARM.
Third, as \ac{HIL} analyses must interact with the physical devices, how is it practicable to consider more than a handful of samples?
C4 is also related since there may be varying ground truth requirements -- why should a code similarity paper limit itself to samples with known bugs when the primary evaluation would greatly benefit from function symbol ground truth instead?

All of these considerations have major impact on the goals G1 and G2, especially considering quality over quantity.
Yet, it appears beneficial to try and attain all other goals for the sake of soundness within the practical limitations of each scenario.

\subsection{Key Corpus Requirements and Their Measures}
\label{sec:reqs:reqs}

Six requirements nurture three goals in~\cref{fig:requirements}.
There are 16 associated measures to assess their fulfillment:
\textit{Date}, \textit{Version}, \textit{Packed}/\textit{Unpacked}, \textit{Link}, \textit{Hash}, \textit{Vulnerabilities}, \textit{Deduplication}, \textit{Reasoning}, \textit{Acquisition}, \textit{Unpacking Process}, \textit{Manufacturer}, \textit{Model}, \textit{Architecture}, \textit{Device Class}, and \textit{FW Type}.
They can feed into multiple requirements.

\begin{itemize}[align=left, wide=0em, leftmargin=1em]
    \item[\textbf{R1}] \textit{Ground Truth.}
    Performance evaluations should search for new and known vulnerabilities.
    The first demonstrates impact and proofs that the method can, indeed, find new bugs.
    The latter helps to verify results and paints scenarios that show what could have been if a tool would have been available in the past.
    As this contributes to representativeness, one should include samples with known \textit{Vulnerabilities}.
    \item[\textbf{R2}] \textit{Relevance.}
    G2 demands samples of relevance.
    In general, we can measure it using temporal properties:
    Samples have a Release \textit{Date} and \textit{Version}.
    These provide information on actuality and history.
    Depending on a paper's scenario, they may vary in effect and meaning for relevance:
    Papers that search for new bugs should consider most recent versions of devices that are still in active use.
    Their end-of-support date could be an indicator for actuality.
    If papers explore, e.g., the proliferation of known bugs, history becomes relevant; then, consider old and new versions.
    Methods may also target certain relevant Device Properties or File Contents (cf. R6).
    \item[\textbf{R3}] \textit{Clean Data.}
    For replicability and sample quality, corpora should contain \textit{Clean Data}.
    Sample \textit{Deduplication} via, e.g., file hashes, helps to clean the results from findings in multiple samples with similar contents.
    We give two examples that can introduce duplicates to the corpus:
    First, scrapers might catch images that are already duplicates on the origin server.
    Second, some manufacturers create base firmware images for whole product lines, which only differ in few files, e.g., drivers.
    Reporting the sample unpacking status also contributes to cleanliness:
    Researchers may provide detail information on the quantities of \textit{Packed} and effective corpus sizes of analyzable and \textit{Unpacked Samples}.
    \item[\textbf{R4}] \textit{Rich Meta Data.}
    Rich meta data helps to ensure replicability within legal possibilities.
    Supplemental data could be file properties like Download \textit{Links}, \textit{Hashes}, \textit{Versions}, and Release \textit{Dates}.
    The latter two also contribute to R2.
    Such information helps to find samples in the Internet when links are broken.
    \textit{Hashes} are, in this regard, disputable when acquisition is (semi-)invasive:
    When we, e.g., dump memory during runtime, it is type-dependent if an image hash is useful.
    ROMs can yield reusable hashes, which could help to verify extraction success.
    But if data changes, the hash changes too, jeopardizing its use.
    Similarity hashes might help in this case.
    Other useful data is the File Contents, e.g., Linux kernel versions, libraries used, or \ac{ISA}.
    Device properties like \textit{Manufacturer} and \textit{Model} can provide further insights on sample distributions.
    Also, meta data serves as proof of relevance and helps third parties to easily extract corpus subsets that nurture their own research goals.
    \item[\textbf{R5}] \textit{Documentation.}
    Providing as much information as possible on sample \textit{Acquisition} and the \textit{Unpacking Process} supports replicability.
    Are there, e.g., any scripts or scrapers?
    If so, can they be shared?
    If not, which steps were followed to obtain samples?
    Which unpackers were used and how did researchers validate their success?
    If unpackers were custom, is it possible to share them?
    It is also useful to describe \textit{Deduplication}.
    Understandable \textit{Reasoning} about sample selection is another aspect of representativeness:
    Which device properties are relevant for the paper and why were the samples chosen?
    Example reasons may be ground truth, availability, heterogeneity, or method limitations.
    \item[\textbf{R6}] \textit{Heterogeneity \& Diversity.}
    Device properties should be as heterogeneous and diverse as possible.
    Example properties are \textit{Manufacturers}, \textit{Models}, \textit{Architectures}, \textit{Firmware Types}, and \textit{Device Classes} such as routers, switches, and network cameras~\cite{costinLargeScaleAnalysisSecurity2014}.
    This helps to draw a good picture of analysis performance across different devices, demonstrates method applicability, and reduces biases introduced by certain properties.
    Applicability, e.g., could be misinterpreted if a corpus only includes manufacturers that do not strip required information like build artifacts~\cite{cve_attribution}.
    Of course, higher heterogeneity implies larger sample quantity, but quality should be preferred:
    There is no benefit in inflating corpus sizes with similar contents or unanalyzable samples.
    Analysis methods ultimately dictate quantities and permutable properties.
    Thus, we may approach heterogeneity with best efforts according to G3 and report \textit{Unpacked Samples}.
\end{itemize}

\section{Analysis: Current Corpus Creation Practices in Top Tier Research}
\label{sec:literature}

We assess if there is common ground on corpus creation practices in research:
We systematically review 44 top tier papers, collect data on our framework, and use the insights to analyze and reveal methodical shortcomings in literature.

\subsection{Paper Selection \& Data Collection Methodology}
\label{sec:literature:methods}

\begin{figure}[t!]
    \begin{center}
        \includegraphics[width=0.925\columnwidth]{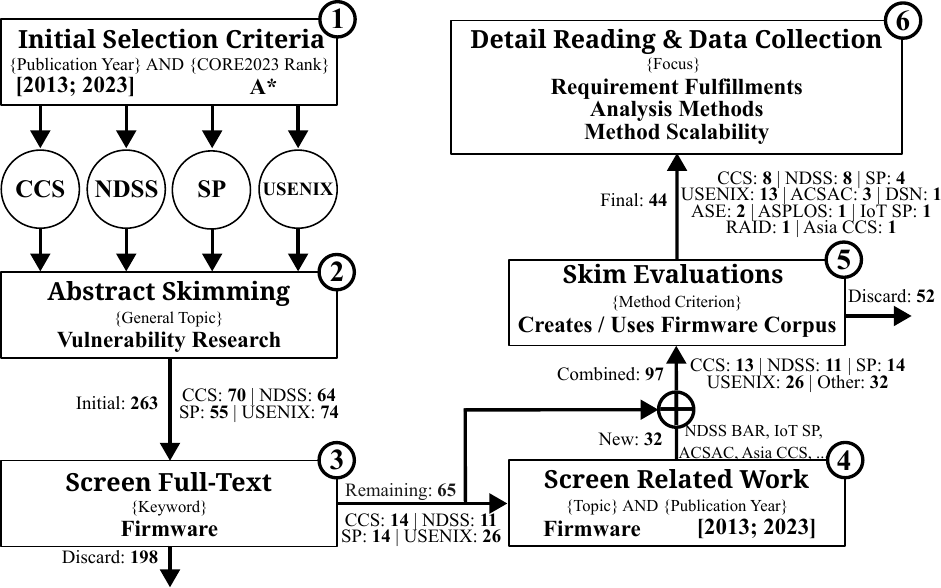}
        \caption{Out of an initial set of 263 peer-reviewed papers from the past ten years, we distilled 44 relevant ones. For each of the remaining papers, we collected data on our 16 corpus measures, which shall support the goals of Replicability and Representativeness (cf.~\cref{sec:reqs}).
        }
        \label{fig:literature:methods}
    \end{center}
\end{figure}

\cref{fig:literature:methods} shows our paper selection process.
We distill top tier papers on vulnerability research that underwent rigorous peer reviews to analyze state of the art scientific practices.

Collection started by downloading all papers from \textit{CCS}, \textit{NDSS}, \textit{SP}, and \textit{USENIX Security}~\circled{1}.
These are the four cybersecurity conferences with the highest rating of A* in the CORE2023.
For actuality, we considered papers from 2013 to 2023.
We skimmed the abstracts and removed all papers that do not focus on vulnerability research~\circled{2}.
The resulting set contained 263 papers.
We then screened their full-text for the keyword \textit{Firmware} and removed items without a match, as they likely do not explore firmware then~\circled{3}.
65 papers remained.
Assuming that high-quality research references other high-quality research, we read the related work sections for referenced work between 2013 and 2023 that focuses on \textit{Firmware} security as well.
We applied the steps~\circled{2} and~\circled{3} to these references, too.
Thus, we effectively dropped the A* requirement and pulled in 32 more papers
from workshops and conferences like \textit{IoT SP}, \textit{ACSAC}, \textit{NDSS BAR}, and \textit{RAID}~\circled{4}.
We skimmed the evaluation methods of the grown set of 97 papers and discarded all papers that do not create or use a firmware corpus~\circled{5}.
The final set, listed in~\cref{tab:literature_list}, has 44 papers from 10 workshops and conferences.
We read every paper and collected data on our requirements using the 16 measures~\circled{6}.
Appendix~\ref{appendix:criteria} provides the catalogue of criteria to assess the fulfillment of all 16 measures.

\begin{table}[t!]
    \caption{Overview of Reviewed Research Papers}
    \label{tab:literature_list}
    \begingroup
    \centering
    \scalebox{0.85375}{
    \begin{tabular}{rccccc}
        \toprule
        \textbf{Paper} & \textbf{Conference} & \textbf{Year} & \textbf{Type} & \textbf{Method} & \textbf{Scal.} \\
        \midrule
        Cui et al.~\cite{cui_when_firmware_attacks}                         & NDSS     & 2013 & S   & P    & \yes        \\
        Costin et al.~\cite{costinLargeScaleAnalysisSecurity2014}           & USENIX Security   & 2014 & S   & P    & \yes        \\
        Avatar~\cite{zaddachAvatarFrameworkSupport2014}                     & NDSS     & 2014 & H & SE;HIL;E   & \no         \\
        Pewny et al.~\cite{pewnyCrossArchitectureBugSearch2015}             & SP       & 2015 & S   & CS         & \unclear    \\
        Firmalice~\cite{shoshitaishviliFirmaliceAutomaticDetection2015}     & NDSS     & 2015 & S   & SE;FA      & \unclear    \\
        PIE~\cite{pie}                                                      & ACSAC    & 2015 & S   & FA;ML      & \unclear    \\
        FIRMADYNE~\cite{chenAutomatedDynamicAnalysis2016}                   & NDSS     & 2016 & D   & E          & \yes        \\
        discovRE~\cite{eschweilerDiscovREEfficientCrossArchitecture2016}    & NDSS     & 2016 & S   & CS         & \yes        \\
        Costin et al.~\cite{costin_dynamic}                                 & Asia CCS      & 2016 & H & P;E  & \yes        \\
        Genius~\cite{genius}                                                & CCS      & 2016 & S   & CS;ML      & \yes        \\
        BootStomp~\cite{bootstomp}                                          & USENIX Security   & 2017 & S   & SE;FA      & \unclear    \\
        FirmUSB~\cite{firmusb}                                              & CCS      & 2017 & S   & SE         & \unclear    \\
        Gemini~\cite{gemini}                                                & CCS      & 2017 & S   & CS;ML      & \yes        \\
        Muench et al.~\cite{muenchWhatYouCorrupt2018}                       & NDSS     & 2018 & H & E;HIL;F    & \no         \\
        DTaint~\cite{dtaint}                                                & DSN      & 2018 & S   & FA         & \yes        \\
        Tian et al.~\cite{attentionspanned}                                 & USENIX Security   & 2018 & S   & P    & \yes        \\
        VulSeeker~\cite{vulseeker}                                          & ASE      & 2018 & S   & CS;ML      & \yes        \\
        FirmUp~\cite{firmup}                                                & ASPLOS   & 2018 & S   & CS         & \yes        \\
        IoTFuzzer~\cite{iotfuzzer}                                          & NDSS     & 2018 & D   & HIL;F      & \no         \\
        FIRM-AFL~\cite{zhengFIRMAFLHighThroughputGreybox2019}               & USENIX Security   & 2019 & D   & E;F        & \yes        \\
        FirmFuzz~\cite{firmfuzz}                                            & IoT SP   & 2019 & H & E;F        & \yes    \\
        SRFuzzer~\cite{srfuzzer}                                            & ACSAC    & 2019 & D   & HIL;F      & \no         \\
        Pretender~\cite{pretender}                                          & RAID     & 2019 & D   & E;HIL      & \no         \\
        HALucinator~\cite{clementsHALucinatorFirmwareRehosting2020}         & USENIX Security   & 2020 & H & E;F        & \unclear    \\
        FirmScope~\cite{elsabaghFIRMSCOPEAutomaticUncovering2020}           & USENIX Security   & 2020 & S   & FA         & \yes        \\
        PDiff~\cite{jiangPDiffSemanticbasedPatch2020}                       & CCS      & 2020 & S   & SA         & \yes        \\
        P$^2$IM~\cite{p2im}                                                 & USENIX Security   & 2020 & H & E;F        & \unclear       \\
        Karonte~\cite{rediniKaronteDetectingInsecure2020}                   & SP       & 2020 & S   & FA         & \yes        \\
        Laelaps~\cite{cao_laelaps}                                          & ACSAC    & 2020 & H & E;SE;F     & \yes        \\
        FirmAE~\cite{firmae}                                                & ACSAC    & 2020 & H & E;F        & \yes        \\
        CPscan~\cite{fuCPscanDetectingBugs2021}                             & CCS      & 2021 & S   & FA         & \yes        \\
        Diane~\cite{diane}                                                  & SP       & 2021 & H & HIL;FA;F   & \no         \\
        DICE~\cite{dice}                                                    & SP       & 2021 & D   & E;F        & \yes        \\
        ECMO~\cite{ecmo_jiang}                                              & CCS      & 2021 & H & E          & \yes        \\
        iFIZZ~\cite{ifizz}                                                  & ASE      & 2021 & H & E;HIL;F    & \no        \\
        Jetset~\cite{johnsonJetsetTargetedFirmware2021}                     & USENIX Security   & 2021 & H & SE;E       & \unclear    \\
        SaTC~\cite{chen_satc}                                               & USENIX Security   & 2021 & S   & FA         & \yes        \\
        Snipuzz~\cite{snipuzz_feng}                                         & CCS      & 2021 & D   & HIL;F   & \no         \\
        $\mu$Emu~\cite{zhouAutomaticFirmwareEmulation2021}                  & USENIX Security   & 2021 & H & SE;E;F     & \yes        \\
        SymLM~\cite{jinSymLMPredictingFunction2022}                         & CCS      & 2022 & S   & ML         & \yes        \\
        Marcelli et al.~\cite{marcelliHowMachineLearning2022}               & USENIX Security   & 2022 & S   & ML  & \yes        \\
        Greenhouse~\cite{greenhouse}                                        & USENIX Security   & 2023 & H & E;FA;F     & \yes        \\
        FirmSolo~\cite{angelakopoulosFirmSoloEnablingDynamic2023}           & USENIX Security & 2023 & H & E;F        & \yes        \\
        VulHawk~\cite{vulhawk}                                              & NDSS     & 2023 & S & CS;ML        & \yes        \\
        \bottomrule
    \end{tabular}
    }
    \endgroup
    \textbf{Scal}ability:
    \yes~= \textit{Scalable};
    \no~= \textit{Not Scalable};
    \unclear~= \textit{Uncertain}.
    \textbf{Method:}
    CS = \textit{Code Similarity};
    E = \textit{Emulation};
    F = \textit{Fuzzing};
    FA = \textit{Flow Analysis};
    HIL = \textit{\acl{HIL}};
    ML = \textit{Machine Learning};
    P = \textit{Pattern};
    SE = \textit{Symbolic Execution}.
    \textbf{Type:}
    S = \textit{Static};
    D = \textit{Dynamic};
    H = \textit{Hybrid}.
\end{table}

\begin{table*}[!t]
    \caption{Corpus Creation Practices in Top Tier Research from 2013 to 2023: Collected Data on the Measures for Scientifically Sound Firmware Corpora}
    \label{tab:literature_results}
    \scalebox{0.98}{
    \begin{tabularx}{\linewidth}{rcccccccccccccccc} 
        \toprule
        
                             & \multicolumn{16}{c}{\textbf{Requirement Applies to Measure}}\\

        \cmidrule[0.4pt](lr){2-17}%

        \textbf{Requirement} &  \rotatebox{78}{\textbf{Packed \#}} & \rotatebox{78}{\textbf{Unpacked \#}}    & \rotatebox{78}{\textbf{Deduplication}} & \rotatebox{78}{\textbf{Unpack Proc.}}     & \rotatebox{78}{\textbf{Reasoning}} & \rotatebox{78}{\textbf{Acquisition}} & \rotatebox{78}{\textbf{Vulnerabilities}} & \rotatebox{78}{\textbf{Rel. Dates}} & \rotatebox{78}{\textbf{Versions}} & \rotatebox{78}{\textbf{Links}} & \rotatebox{78}{\textbf{Hashes}} & \rotatebox{78}{\textbf{Manufacturer}} & \rotatebox{78}{\textbf{Models}} & \rotatebox{78}{\textbf{Dev. Classes}} & \rotatebox{78}{\textbf{ISAs}}  & \rotatebox{78}{\textbf{FW Types}}\\
        
        \cmidrule[0.4pt](lr){1-1}%
        \cmidrule[0.4pt](lr){2-2}%
        \cmidrule[0.4pt](lr){3-3}%
        \cmidrule[0.4pt](lr){4-4}%
        \cmidrule[0.4pt](lr){5-5}%
        \cmidrule[0.4pt](lr){6-6}%
        \cmidrule[0.4pt](lr){7-7}%
        \cmidrule[0.4pt](lr){8-8}%
        \cmidrule[0.4pt](lr){9-9}%
        \cmidrule[0.4pt](lr){10-10}%
        \cmidrule[0.4pt](lr){11-11}%
        \cmidrule[0.4pt](lr){12-12}%
        \cmidrule[0.4pt](lr){13-13}%
        \cmidrule[0.4pt](lr){14-14}%
        \cmidrule[0.4pt](lr){15-15}%
        \cmidrule[0.4pt](lr){16-16}%
        \cmidrule[0.4pt](lr){17-17}%

        \textbf{R1)} Ground Truth     & --             & --             & --             & --             & --             & --             & \CheckmarkBold & --             & --             & --             & --             & --             & --             & --             & --             & -- \\ 
        \textbf{R2)} Relevance        & --             & --             & --             & --             & --             & --             & --             & \CheckmarkBold & \CheckmarkBold & --             & --             & \CheckmarkBold & \CheckmarkBold & \CheckmarkBold & \CheckmarkBold & \CheckmarkBold  \\ 
        \textbf{R3)} Clean Data       & \CheckmarkBold & \CheckmarkBold & \CheckmarkBold & --             & --             & --             & --             & --             & --             & --             & --             & --             & --             & --             & --             & -- \\ 
        \textbf{R4)} Rich Meta Data   & --             & --             & --             & --             & --             & --             & --             & \CheckmarkBold &\CheckmarkBold  & \CheckmarkBold & \CheckmarkBold & \CheckmarkBold & \CheckmarkBold & \CheckmarkBold & \CheckmarkBold & \CheckmarkBold  \\
        \textbf{R5)} Documentation    & --             & --             & \CheckmarkBold & \CheckmarkBold & \CheckmarkBold & \CheckmarkBold & --             & --             & --             & --             & --             & --             & --              & --             & --            & --\\
        \textbf{R6)} Heterogeneity    & --             & \CheckmarkBold & --             & --             & --             & --             & --             & --             & --             & --             & --             & \CheckmarkBold & \CheckmarkBold & \CheckmarkBold & \CheckmarkBold & \CheckmarkBold \\

        \midrule
        
        \textbf{Paper} & \multicolumn{16}{c}{\textbf{Collected Data on the Measures for Scientifically Sound Fimware Corpora}}\\
        
        \cmidrule[0.4pt](lr){1-1}%
        \cmidrule[0.4pt](lr){2-17}%

        Cui et al.~\cite{cui_when_firmware_attacks}                         & 373                     & \no                         & \no      & \yes                 & \yes                      & \no                   & \yes                     & \yes                      & \unclear                 & \no                   & \no                        & 1                       & 63                     & 1                            & 2                   & II                      \\
        Costin et al.~\cite{costinLargeScaleAnalysisSecurity2014}           & 32,356                  & 26,275                      & \no      & \yes                 & \unclear                  & S                     & \unclear                 & \no                       & \no                      & \yes                  & \yes                       & \unclear                & \unclear               & \unclear                     & \unclear            & \unclear                \\
        Avatar~\cite{zaddachAvatarFrameworkSupport2014}                     & 3                       & 3                           & \yes     & \no                  & \yes                      & M                     & \unclear                 & \no                       & \no                      & \no                   & \no                        & 3                       & 3                      & 3                            & 1                   & II-III                  \\
        Pewny et al.~\cite{pewnyCrossArchitectureBugSearch2015}             & 6                       & 6                           & \yes     & \no                  & \yes                      & M                     & \yes                     & \no                       & \yes                     & \yes                  & \no                        & 6                       & 6                      & 3                            & 3                   & 0-I                     \\
        PIE~\cite{pie}                                                      & 4                       & 4                           & \yes     & \no                  & \unclear                  & \no                   & \no                      & \no                       & \no                      & \no                   & \no                        & \unclear                & 4                      & 4                            & 1                   & III                     \\
        Firmalice~\cite{shoshitaishviliFirmaliceAutomaticDetection2015}     & 3                       & 3                           & \yes     & \no                  & \yes                      & M                     & \yes                     & \no                       & \no                      & \no                   & \no                        & 3                       & 3                      & 3                            & 2                   & I                       \\
        FIRMADYNE~\cite{chenAutomatedDynamicAnalysis2016}                   & 23,035                  & 9,486                       & \yes     & \yes                 & \unclear                  & S                     & \yes                     & \yes                      & \yes                     & \yes                  & \yes                       & 42                      & \unclear               & \unclear                     & 7                   & I-II                    \\
        discovRE~\cite{eschweilerDiscovREEfficientCrossArchitecture2016}    & 3                       & 3                           & \yes     & \no                  & \yes                      & M                     & \yes                     & \no                       & \yes                     & \yes                  & \yes                       & 3                       & 3                      & 3                            & 4                   & 0-I                     \\
        Costin et al.~\cite{costin_dynamic}                                 & 1,925                   & 1,925                         & \no      & \no                  & \yes                      & \no                   & \unclear                 & \no                       & \unclear                 & \no                   & \no                        & \unclear                & \unclear               & \unclear                     & 9                   & I                       \\
        Genius~\cite{genius}                                                & 33,045                  & 8,126                       & \no      & \no                  & \no                       & S;R                   & \unclear                 & \no                       & \no                      & \yes                  & \no                        & 26                      & \unclear               & \unclear                     & \unclear            & \unclear                \\
        BootStomp~\cite{bootstomp}                                          & 5                       & 5                           & \yes     & \unclear             & \yes                      & M                     & \yes                     & \no                       & \unclear                 & \no                   & \no                        & 4                       & 4                      & 1                            & 1                   & III                     \\
        FirmUSB~\cite{firmusb}                                              & 2                       & 2                           & \yes     & \unclear             & \no                       & M                     & \no                      & \no                       & \no                      & \no                   & \no                        & 2                       & 2                      & 1                            & \unclear            & III                     \\
        Gemini~\cite{gemini}                                                & 33,045                  & 8,126                       & \no      & \no                  & \no                       & R                     & \unclear                 & \no                       & \no                      & \yes                  & \no                        & 26                      & \unclear               & \unclear                     & \unclear            & \unclear                \\
        Muench et al.~\cite{muenchWhatYouCorrupt2018}                       & 4                       & 4                           & \yes     & \no                  & \yes                      & M                     & \unclear                 & \no                       & \no                      & \no                   & \no                        & 4                       & 4                      & 4                            & 1                   & 0-III                   \\
        DTaint~\cite{dtaint}                                                & 6                       & 6                           & \yes     & \no                  & \yes                      & \no                   & \yes                     & \no                       & \yes                     & \no                   & \no                        & 4                       & 6                      & \unclear                     & 2                   & I                       \\
        Tian et al.~\cite{attentionspanned}                                 & 2,018                   & \unclear                    & \no      & \yes                 & \yes                      & S                     & \notapplicable           & \no                       & \yes                     & \unclear              & \no                        & 11                      & \unclear               & 1                            & \notapplicable      & I                       \\
        VulSeeker~\cite{vulseeker}                                          & 4,643                   & \no                         & \no      & \unclear             & \no                       & R                     & \unclear                 & \no                       & \no                      & \yes                  & \no                        & \unclear                & \unclear               & \unclear                     & \unclear            & \unclear                \\
        FirmUp~\cite{firmup}                                                & \unclear 5,000          & \unclear 2,000              & \no      & \yes                 & \no                       & S                     & \yes                     & \no                       & \no                      & \no                   & \no                        & \unclear                & \unclear               & \unclear                     & \unclear            & \unclear                \\
        IoTFuzzer~\cite{iotfuzzer}                                          & 17                      & \notapplicable              & \yes     & \notapplicable       & \yes                      & \no                   & \yes                     & \no                       & \yes                     & \no                   & \no                        & 12                      & 17                     & 10                           & \unclear            & \unclear                \\
        FIRM-AFL~\cite{zhengFIRMAFLHighThroughputGreybox2019}               & 11                      & 11                          & \yes     & \no                  & \no                       & M;R                   & \yes                     & \no                       & \yes                     & \no                   & \no                        & 5                       & 11                     & 2                            & \unclear            & I                       \\
        FirmFuzz~\cite{firmfuzz}                                            & 6,427                   & 1,013                       & \yes     & \no                  & \yes                      & S                     & \yes                     & \no                       & \no                      & \no                   & \no                        & 3                       & \unclear               & 1                            & 2                   & I                       \\
        SRFuzzer~\cite{srfuzzer}                                            & 10                      & \notapplicable              & \yes     & \notapplicable       & \no                       & M                     & \no                      & \no                       & \yes                     & \no                   & \no                        & 5                       & 10                     & 1                            & 2                   & \unclear                \\
        Pretender~\cite{pretender}                                          & 6                       & \notapplicable              & \yes     & \notapplicable       & \unclear                  & M                     & \no                      & \no                       & \no                      & \yes                  & \yes                       & 2                       & 3                      & 1                            & 1                   & III                     \\
        HALucinator~\cite{clementsHALucinatorFirmwareRehosting2020}         & 16                      & 16                          & \yes     & \no                  & \yes                      & M                     & \unclear                 & \no                       & \no                      & \yes                  & \no                        & 3                       & 4                      & 1                            & 1                   & III                     \\
        FirmScope~\cite{elsabaghFIRMSCOPEAutomaticUncovering2020}           & 2,017                   & \unclear                    & \no      & \yes                 & \unclear                  & S                     & \yes                     & \no                       & \no                      & \unclear              & \no                        & 99+                     & \unclear               & 1                            & \notapplicable      & I                       \\
        PDiff~\cite{jiangPDiffSemanticbasedPatch2020}                       & 715                     & \no                         & \no      & \no                  & \no                       & \no                   & \yes                     & \no                       & \no                      & \no                   & \no                        & 8                       & \unclear               & 3                            & 2                   & I                       \\
        P$^2$IM~\cite{p2im}                                                 & 10                      & 10                          & \yes     & \unclear             & \yes                      & M                     & \no                      & \no                       & \no                      & \yes                  & \no                        & 3                       & 4                      & 10                           & 1                   & II-III                  \\
        Karonte~\cite{rediniKaronteDetectingInsecure2020}                   & 53;899                  & \unclear                    & \yes     & \yes                 & \yes                      & S;R                   & \yes                     & \yes                      & \yes                     & \yes                  & \yes                       & 25                      & \unclear               & \unclear                     & 3                   & I-III                   \\
        Laelaps~\cite{cao_laelaps}                                          & 30                      & \notapplicable              & \yes     & \unclear             & \yes                      & \no                   & \no                      & \no                       & \no                      & \no                   & \no                        & 2                       & 4                      & 24                           & 1                   & II-III                  \\
        FirmAE~\cite{firmae}                                                & 1,306                   & 1,124                       & \yes     & \yes                 & \yes                      & S                     & \yes                     & \yes                      & \no                      & \yes                  & \yes                       & 8                       & \unclear               & 2                            & 2                   & I                       \\
        CPscan~\cite{fuCPscanDetectingBugs2021}                             & 28                      & 28                          & \yes     & \no                  & \no                       & \no                   & \no                      & \no                       & \yes                     & \no                   & \no                        & 10                      & 28                     & \unclear                     & \unclear            & I                       \\
        Diane~\cite{diane}                                                  & 11                      & \notapplicable              & \yes     & \notapplicable       & \no                       & \no                   & \yes                     & \no                       & \yes                     & \no                   & \no                        & 9                       & 11                     & 4                            & \unclear            & \unclear                \\
        DICE~\cite{dice}                                                    & 7                       & \notapplicable              & \yes     & \notapplicable       & \yes                      & M                     & \no                      & \no                       & \no                      & \yes                  & \no                        & 6                       & 7                      & 7                            & 1                   & II-III                  \\
        ECMO~\cite{ecmo_jiang}                                              & 815                     & 815                         & \no      & \yes                 & \unclear                  & \no                   & \no                      & \no                       & \no                      & \no                   & \no                        & 2                       & 37                     & 1                            & 1                   & I                       \\
        iFIZZ~\cite{ifizz}                                                  & 10                      & 10                          & \yes     & \yes                 & \yes                      & \no                   & \unclear                 & \no                       & \yes                     & \no                   & \no                        & 7                       & 10                     & 4                            & 2                   & I                       \\
        Jetset~\cite{johnsonJetsetTargetedFirmware2021}                     & 13                      & 13                          & \yes     & \no                  & \unclear                  & M;R                   & \no                      & \no                       & \no                      & \no                   & \no                        & 4                       & 13                     & 3                            & 3                   & I-III                   \\
        SaTC~\cite{chen_satc}                                               & 39;49                   & 39;49                       & \yes     & \yes                 & \no                       & \no;R                 & \no                      & \no                       & \yes                     & \yes                  & \yes                       & 6;4                     & 6;\unclear             & 2;\unclear                   & 2;3                 & \unclear                \\
        Snipuzz~\cite{snipuzz_feng}                                         & 20                      & \notapplicable              & \no      & \notapplicable       & \yes                      & M                     & \no                      & \no                       & \yes                     & \no                   & \no                        & 17                      & 20                     & 8                            & \unclear            & \unclear                \\
        $\mu$Emu~\cite{zhouAutomaticFirmwareEmulation2021}                  & 21                      & 21                          & \yes     & \no                  & \no                       & M;R                   & \yes                     & \no                       & \yes                     & \yes                  & \no                        & \unclear                & 21                     & \unclear                     & 1                   & II-III                  \\
        SymLM~\cite{jinSymLMPredictingFunction2022}                         & 8                       & 8                           & \yes     & \no                  & \unclear                  & R                     & \notapplicable           & \no                       & \no                      & \no                   & \no                        & \unclear                & 8                      & \unclear                     & 1                   & II-III                  \\
        Marcelli et al.~\cite{marcelliHowMachineLearning2022}               & 2                       & 2                           & \yes     & \no                  & \yes                      & M                     & \yes                     & \no                       & \no                      & \no                   & \no                        & 2                       & 2                      & 1                            & 2                   & I                       \\
        Greenhouse~\cite{greenhouse}                                        & 7,141                   & 5,690                       & \yes     & \yes                 & \yes                      & S;R                   & \yes                     & \no                       & \no                      & \no                   & \no                        & 9                       & 1,764                  & 2                            & 3                   & I                       \\
        FirmSolo~\cite{angelakopoulosFirmSoloEnablingDynamic2023}           & 8,737                   & 1,470                       & \yes     & \unclear             & \yes                      & \no;R                 & \yes                     & \no                       & \unclear                 & \unclear              & \no                        & \unclear                & \unclear               & \unclear                     & 2                   & I                       \\
        VulHawk~\cite{vulhawk}                                              & 20                      & 20                          & \no      & \no                  & \no                       & \no                   & \yes                     & \no                       & \no                      & \no                   & \no                        & 3                       & 20                     & \unclear                     & \unclear            & \unclear                \\
        \midrule \midrule
        LFwC (\cref{sec:corpus})                                                         & 14,583                  & 10,913                      & \yes     & \yes                 & \yes                      & S                     & \yes                     & \yes                      & \yes                     & \yes                  & \yes                       & 10                      & 2,365                  & 22         & 9                   & I-II                    \\
        \bottomrule
    \end{tabularx}
    }
    \textbf{Semicolon (;):} Multiple Methods/Corpora/Data Points.
    \textbf{Symbols:} \CheckmarkBold = \textit{Requirement Applies to Data Column Below}; \yes~= \textit{Documented/Proof of Presence in Data}; \no~= \textit{Undocumented/Proof of Absence in Data}; \unclear~= \textit{Partially Documented/Missing Data to Proof Absence or Presence}; \notapplicable~= \textit{Not Applicable in Paper Scenario}.
    \textbf{Acquisition:} S = \textit{Web-Scraping}; M = \textit{Manual Collection}; R = \textit{Samples from Related Work}. \textbf{Firmware Types:} As described in Appendix~\ref{appendix:types}.
\end{table*}

\subsection{General Statistics \& Result Overview}
\label{sec:literature:stats}
\cref{tab:literature_list} shows the papers and gives insights on method types, i.e., static, dynamic, or hybrid.
Publication year, conference, scalability, and analysis method are noted as well.

Each year from 2013 to 2023 is represented, with rising quantities until 2021.
Fewer papers were published in 2022 and 2023.
The four most represented conferences are \textit{USENIX Security} (13 papers), \textit{CCS} (8), \textit{NDSS} (8), and \textit{SP} (4).
22 papers use static, seven dynamic, and 15 hybrid approaches.

28 methods are rated as scalable (\yes).
FIRMADYNE~\cite{chenAutomatedDynamicAnalysis2016}, e.g., analyzes 9,500 samples.
We rated seven papers as unscalable (\no); all apply \ac{HIL}.
For nine papers, there was uncertainty regarding scalability (\unclear).
HALucinator~\cite{clementsHALucinatorFirmwareRehosting2020} and P$^2$IM~\cite{p2im}, e.g., both
use emulation but need manual modelling.
Their corpora are small and do not reveal scalability.
Modelling efforts can not be estimated without in-depth system experience.

\cref{tab:literature_results} provides the results on all measures in our framework.
The upper half maps the 16 measures to their associated requirements from~\cref{sec:reqs:reqs} with a check mark (\CheckmarkBold).
The lower half provides the collected data for each paper.
Where possible, we provide concrete values from the papers.
If not, we resort to a symbolism marking complete~(\yes), partial~(\unclear), or missing~(\no) documentation.
When a measure is not applicable to a specific scenario, we insert the \notapplicable~symbol.

The data allows us to study various practices in firmware corpus creation.
We start by sharing preliminary observations we made while working with the raw data in~\cref{tab:literature_results}.
Then, we group the results by measure to study their overall performance across all papers.
Next, we discuss selected details on corpus creation practices with focus on the impact of the quality over quantity credo from~\cref{sec:reqs:goals}.
Finally, we group findings by requirement and conclude upon their fulfillment.

\subsection{Preliminary Observations}
\label{sec:literature:general}

$\blacktriangleright$ \textbf{A paper's scenario dictates feasible quantities.}
G3 says that all numbers in~\cref{tab:literature_results} are relative to their paper scenario.
Aspects like sample accessibility and scalability influence experiments.
The data backs this claim:
\cref{tab:literature_results} reveals that 17 out of 44 papers use corpora between 373 and 33,000 samples.
All of them use scalable methods according to~\cref{tab:literature_list}; the majority of them scrapes the accessible Type-I.
Only one of the 17 papers includes the specialized and harder to acquire Type-III.
Vice versa, 27 papers use corpora of two to 49 samples.
Out of these, 70\% either target Type-III or use \ac{HIL}.
Low quantities must not mean bad practice, as they can also reveal limits of feasibility in spite of best efforts.
This does not change the fact that they introduce statistical uncertainty.

$\blacktriangleright$ \textbf{The measures are practicable and relevant.}
We have created a framework that addresses scientifically relevant aspects of corpus creation.
Thus, we proposed concrete measures to test requirement fulfillment (cf.~\cref{sec:reqs:reqs}).
We aimed to select measures with research relevance and universal applicability.
\cref{tab:literature_results} holds 704 data points across the 16 measures and 44 papers.
We considered that measures may not be applicable (\notapplicable), too.
This is only true for 17 out of 704 data points:
The high viability of all measures in literature let us conclude that our framework is practicable and can find broad application in research.

\subsection{Quantitative Result Analysis by Measure}
\label{sec:literature:by_measure}

\begin{figure}[t]
    \begin{center}
        \includegraphics[width=0.9\columnwidth]{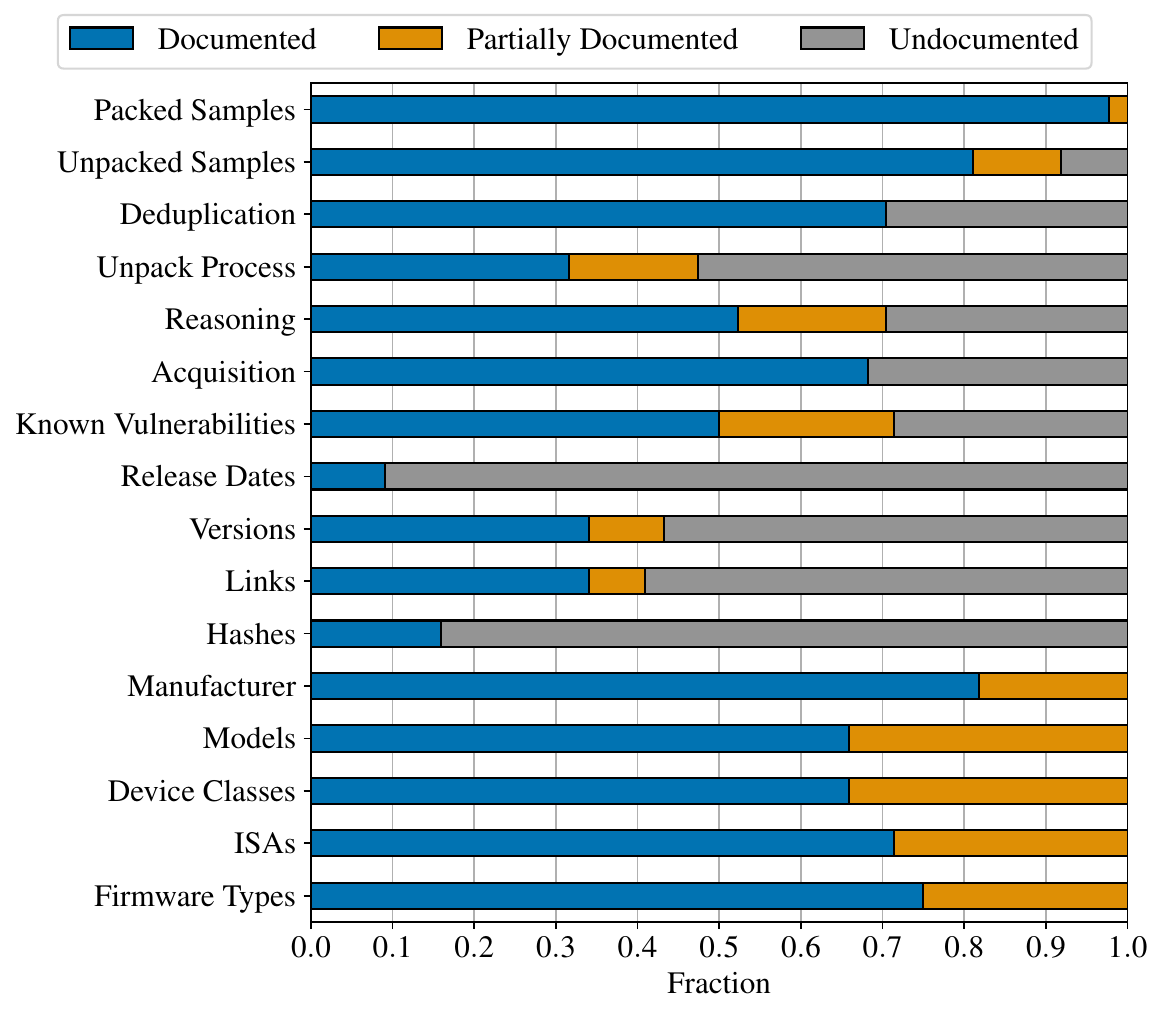}
        \caption{Aggregated results of all collected data points for each measure in~\cref{tab:literature_results}. Data points that mark non-applicability of a measure are considered.}
        \label{fig:literature:measure_docs}
    \end{center}
\end{figure}

We quantitatively analyze the data in~\cref{tab:literature_results} to identify the cumulative measure performance across all papers and discuss current practices in research.
We group data by measure and convert concrete numbers to the value~\yes.
Thus, we establish a comparison baseline using four discrete values:
A paper documents the subject of a measure fully~(\yes), partially~(\unclear), or not at all~(\no).
The fourth is non-applicability~(\notapplicable).
We calculated the fraction of documented data points for a measure across all applicable papers.
\cref{fig:literature:measure_docs} shows the unweighted results.

$\blacktriangleright$ \textbf{[Sample Quantities] All document packed samples. Some omit unpacked quantities.}
All papers precisely specify the quantity of packed samples.
FirmUp~\cite{firmup}, however, does not.
The performance drops to 91\% for unpacked samples:
36 papers (81\%) provide precise unpacking numbers while four papers only give partial information (10\%)~\cite{firmup,rediniKaronteDetectingInsecure2020,attentionspanned,elsabaghFIRMSCOPEAutomaticUncovering2020}:
One gives approximations, three other include unpacking as system component but do not provide clear numbers.
Three papers do not share any unpacked quantities~\cite{cui_when_firmware_attacks,vulseeker,jiangPDiffSemanticbasedPatch2020}.

$\blacktriangleright$ \textbf{[Deduplication] 30\% of the papers do not describe sample deduplication.}
As we will discuss in detail in~\cref{sec:literature:key_observations}, sample deduplication is important to avoid skewness in analysis results due to, e.g., duplicate findings.
We note that the performance on this measure is over-evaluated in terms of documentation awareness:
The 70\% already includes papers that share artifacts, which helped us to determine if any deduplication took place (cf.~Appendix~\ref{appendix:criteria}).

$\blacktriangleright$ \textbf{[Unpacking Process] 52\% of the papers do not describe the unpacking process.}
Sample unpacking is a significant barrier to any kind of replicability and, thus, result verification.
20 (52\%) of all papers that are applicable to this measure do not document the critical unpacking process.
12 (32\%) document it in detail, e.g, Greenhouse~\cite{greenhouse}, FirmScope~\cite{elsabaghFIRMSCOPEAutomaticUncovering2020}, and Karonte~\cite{rediniKaronteDetectingInsecure2020}.
Six (16\%) document it partially.

$\blacktriangleright$ \textbf{[Reasoning] 13 papers do not justify sample selection.}
It is useful for third parties to understand \textit{why} a corpus contains certain samples, as such information gives insights on possible limitations and goals.
It further helps to contextualize the work and interpret results.
Possible reasons for sample selection could be, e.g., availability, required firmware properties like \acp{ISA}, or a device class of particular interest.
30\% of papers do not give a reason, 18\% give a reason that was not entirely comprehensible to us, and 52\% justify comprehensively.

$\blacktriangleright$ \textbf{[Acquisition] 32\% of the papers do not document acquisition.}
Sharing \textit{how} samples were acquired points independent research into the direction of corpus replication, be it through scraping or manual firmware extraction.
14 out of 44 (32\%) papers do not provide any information on this matter.

$\blacktriangleright$ \textbf{[Known Vulnerabilities] 50\% of the papers have no or incomplete documentation on the existence of known bugs in their corpora.}
The existence of known vulnerabilities in corpora helps to obtain verifiable evidence showing the fruitfulness of a new analysis method.
\textit{If} there are known bugs fitting to the paper, it is a choice to use them as benchmark.
21 out of 42 papers fully document the existence of ground truth (50\%).
DTaint~\cite{dtaint}, e.g., rediscovers six verifiable CVEs in their corpus.
Nine papers partially document this subject (21\%).
VulSeeker~\cite{vulseeker}, e.g., searches for CVE-2015-1791, but does not explain which samples are affected in the corpus.
Experiments using other CVEs are mentioned, but also not explained.
12 papers do not mention ground truth~(29\%).

$\blacktriangleright$ \textbf{[File \& Temporal Properties] Release dates, versions, links, and hashes are rarely documented.}
Considering temporal properties that could help to estimate relevance, only four out of 44 papers report firmware release dates (4\%) and 15 (34\%) report firmware versions.
For the latter, there are four more papers with partial data:
BootStomp~\cite{bootstomp}, e.g., reports experiments on an \textit{older} and \textit{newer} bootloader version by Qualcomm, but does not name the identifiers.
File properties beneficial to replicability are also rarely documented:
15 out of 44 papers share links for download or device acquisition.
If such links become invalid, readers can fall back to file hashes to find alternative sources.
Three papers provide an incomplete sample list, e.g., FirmSolo~\cite{angelakopoulosFirmSoloEnablingDynamic2023}, who use the fully documented FIRMADYNE~\cite{chenAutomatedDynamicAnalysis2016} corpus but then add 50 samples of unknown origin.
Hashes are available in seven out of 44 cases.

$\blacktriangleright$ \textbf{[Device Properties] All papers discuss corpus composition regarding heterogeneous device properties.}
In all papers, there is full or partial information on the device properties Manufacturer, Model, Device Class, \ac{ISA}, and Firmware Type.
This is a positive result as it shows that all papers provide insights on heterogeneity.
Yet, between 25\% and 34\% of papers only give partial information.
They can be grouped into two classes:
First, there are papers that bulk scrape images but do not collect meta data, e.g., Costin et al.~\cite{costinLargeScaleAnalysisSecurity2014}.
Second, some papers give incomplete information on these properties.
FirmUp~\cite{firmup}, e.g., lists example manufacturers, but not all of them.
In both cases, some device properties remain unknown, which makes it harder to assess corpus composition.

\subsection{On Quality over Quantity for Representative Results}
\label{sec:literature:key_observations}

Our requirements are extensive and strict.
As we will see in~\cref{sec:corpus}, creating a corpus that caters to all included measures is difficult.
It demands attention to aspects that are usually \textit{not} core paper contributions.
Furthermore, documentation occupies space that is a valuable asset in papers.

It is understandable and of little surprise that we found qualitative issues in corpus creation when we gathered the data for~\cref{tab:literature_results}.
We found that discussions on common practices that are \textit{not} covered by our measures provide valuable insights on the importance of the quality over quantity credo from G2:

$\blacktriangleright$ \textbf{Put special attention to packed and unpacked sample quantities while writing and reading.}
Most papers provide all sample quantities we searched for.
But we often caught ourselves revising the collected data while reading through papers, especially with large corpora.
Authors should put special attention to clearly communicate corpus statistics, as imprecise wording can unintentionally skew perspectives on representativeness.
Two arbitrarily chosen examples from otherwise excellent work show how easy it is to fall into this trap:
Costin et al.~\cite{costinLargeScaleAnalysisSecurity2014} report in the abstract that they \textit{unpacked} 32,000 images.
Yet, later they describe 32,356 \textit{processed files} and 26,275 \textit{successfully unpacked} images; the net corpus size shrinks by 6,000.
Genius~\cite{genius} combines two corpora from related work~\cite{costinLargeScaleAnalysisSecurity2014,chenAutomatedDynamicAnalysis2016}.
The abstract reports evaluations on a \textit{data set} of 33,045 \textit{devices}.
Later statements reveal that 8,126 \textit{firmware images were successfully unpacked} for evaluation.
There is 76\% loss.
The former puts perspective on the importance of high-quality, analyzable samples.
As for the latter, peeking into one~\cite{chenAutomatedDynamicAnalysis2016} of the corpora reveals that multiple images exist for one device, showing that interchanging the terms \textit{devices} and \textit{firmware images} leads to ambiguity.

$\blacktriangleright$ \textbf{Bulk collection, combinations, and deduplication.}
Data uniqueness serves sample quality, as duplicates lead to skewed analysis results.
The fact that we did not find deduplication information in ten out of 17 papers with corpora of 373 or more samples is worrying, as bulk collection may catch the same sample multiple times.
Thus, we can not verify that deduplication took place.
The practice of combining multiple corpora from related work that scraped partially overlapping sources feeds into this issue, e.g., Feng et al.~\cite{genius}.

$\blacktriangleright$ \textbf{Contents: Open source samples inflate corpora.}
OpenWrt~\cite{openwrt} and DD-WRT~\cite{ddwrt} are two open source distributions that provide an alternative \ac{OS} for consumer-grade network devices.
Their container formats are well-known and can be easily unpacked.
Thus, it is understandable that we observe the inclusion of such highly available and analyzable samples into corpora to evade the unpacking barrier.
Yet, their disproportional inclusion inflates firmware corpora with content-based duplicates and skews heterogeneity.
These projects share a common code base across devices, \acp{ISA}, and build settings.
We compared the contents of two factory images of OpenWRT v23.05.0, released on 2023-10-12.
We selected builds for the ASUS RT-AC87U~\cite{openwrt_asus} and NETGEAR R8000~\cite{openwrt_netgear} consumer routers and used FACT v4.2-dev~\cite{fact} for unpacking.
SHA256 comparisons yield that the included files overlap by 95\%.
Angelakopoulos et al.~\cite{angelakopoulosFirmSoloEnablingDynamic2023} also see this problem and point out that they removed 4,020 DD-WRT and OpenWrt samples from the 9,486 successfully unpacked samples of FIRMADYNE~\cite{chenAutomatedDynamicAnalysis2016}.
These samples represent 42\% of the original evaluation corpus.
Removing them impacts representativeness.

\subsection{Are Current Practices Meeting our Requirements?}
\label{sec:literature:conclusions}

Is current research meeting our requirements?
Like~\cref{fig:literature:measure_docs}, we aggregate the collected data points from~\cref{tab:literature_results} across all applicable papers, but group them by requirement instead.
We calculate the share of full (\yes), partial (\unclear), and missing (\unclear) documentation per requirement.
Data is unweighted \textit{within} a single requirement, but weighted \textit{across} requirements, as some measures contribute to multiple requirements (cf.~\cref{tab:literature_results}).

\begin{figure}[t]
    \begin{center}
        \includegraphics[width=0.95\columnwidth]{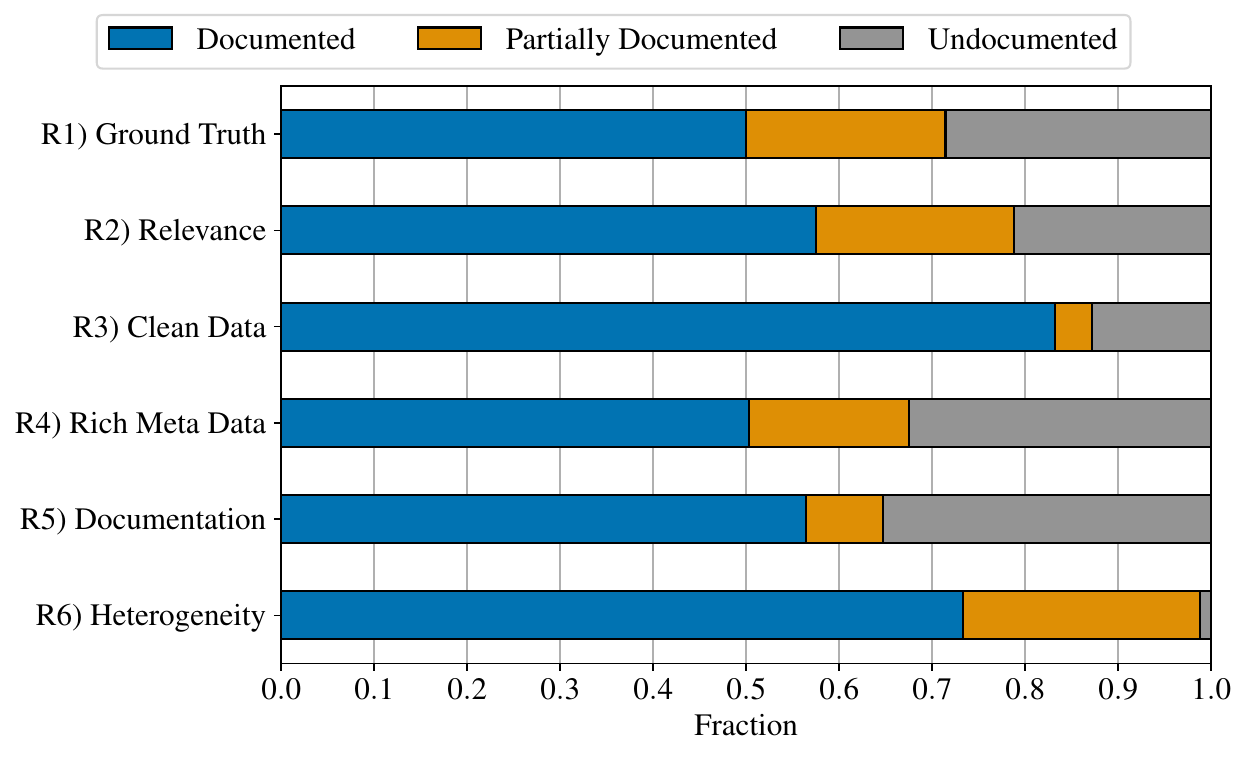}
        \caption{Aggregates the results of all collected data points for the associated measures in~\cref{tab:literature_results}. The associated measures are unweighted \textit{within} a requirement, but weighted \textit{across} requirements, because they can contribute towards multiple goals. All 44 papers are included and data points that mark non-applicability of a measure are considered.}
        \label{fig:literature:req_results}
    \end{center}
\end{figure}

\cref{fig:literature:req_results} shows that researchers put effort into corpus creation.
Yet, there is room for improvement:
They could include more meta data for better replicability and representativeness (R4).
One should provide release dates, versions, download links, and file~hashes~(R2, R4).
Subjects covered by R5 should be thoroughly documented.
Especially unpacking steps often remain unclear.
Regarding our observations on the impact of the quality over quantity credo~(cf.~\cref{sec:literature:key_observations}), we argue that there are many step stones such as missing deduplication that must be documented to draw a better picture on representativeness and provide clean data for R3.
Researchers may conduct more experiments that search for known vulnerabilities in firmware (R1).
Finally, it is wise to improve the precision on all aspects of R6 -- through documentation or artifact sharing.

Thus, current practices in firmware vulnerability research meet our requirements only partially:
None of the 44 reviewed papers documents the subject of all 16 measures.
The results of this literature analysis show that there is currently no common ground on sound firmware corpus creation and documentation.
Missing meta data, incomplete documentation, and inflated corpora blur visions on representativeness and replicability.

Generally, we see that otherwise excellent work may fall into the trap of the methodological and practical challenges we discussed in~\cref{sec:challenges}.
A brief analysis, in which we re-clustered the results from~\cref{fig:literature:req_results} by publication year, did not reveal any rising or declining trends of documentation awareness in this research branch.

\begin{figure*}[t!]
    \begin{center}
        \includegraphics[width=0.925\textwidth]{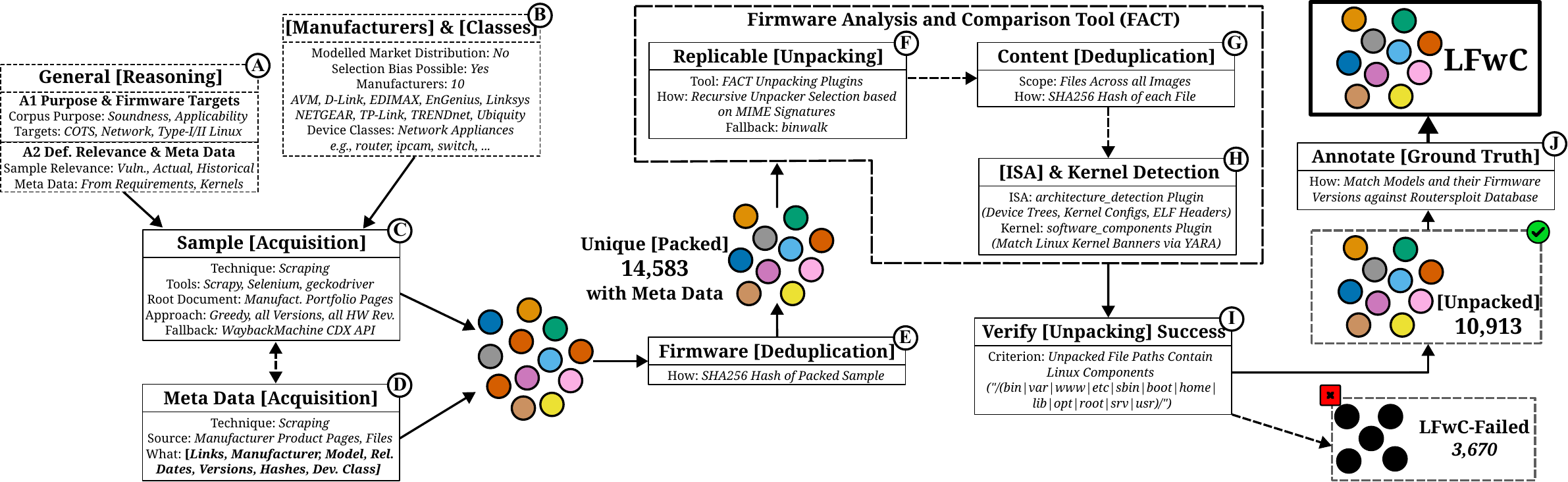}
        \caption{We summarize and document the \ac{LFwC} corpus creation process in a flow graph. There are ten processing steps, identified by the letters A-J. We provide a process revolving around the Firmware Analysis and Comparison Tool (FACT)~\cite{fact} for replicable unpacking success, content deduplication, and analyses for additional meta data. Addressed measures from our Requirements Framework in~\cref{sec:reqs} are marked by [Square Brackets].}
        \label{fig:corpus_creation}
    \end{center}
\end{figure*} 

\section{LFwC: A New Corpus to Demonstrate the Practicability of The Proposed Requirements}
\label{sec:corpus}

We built a proof of concept \acl{LFwC}~(\acs{LFwC}) to assess the feasibility of our requirements.
It is based on data until June 2023 and consists of 10,913 deduplicated and unpacked firmware images from ten known manufacturers.
It includes recent and historical firmware, covering 2,365 unique devices across 22 classes.
To provide an overview of \ac{LFwC}, we added corpus data points to the bottom of~\cref{tab:literature_results}.

We share plenty of meta data and publish all scripts and tools for replicability.
We tear down \ac{LFwC}'s unpacking barrier with an open source process.
Access to the meta data can be requested on Zenodo~\cite{lfwc_full_zenodo}.
The tools and artifacts are available at \url{https://github.com/fkie-cad/linux-firmware-corpus}.

\subsection{Corpus Creation}
\label{sec:corpus:creation}

\cref{fig:corpus_creation} shows the corpus creation process explained below.

\noindent \circled{A} \textbf{General Reasoning.} We formulate two statements:
\begin{itemize}[align=left, wide=0em, leftmargin=1em]
    \item[\textbf{A1}] \textit{Purpose \& Firmware Targets.}
    \ac{LFwC} aims to add value for vulnerability research while being as sound as possible.
    It shall cover multiple paper scenarios with a sizable quantity of images.
    We target Linux firmware, as it is prevalent in research.
    More precisely, we target \ac{COTS} network appliances due to their availability.
    \item[\textbf{A2}] \textit{Definition of Sample Relevance \& Meta Data Reasoning.}
    Samples with vulnerability ground truth, but also historical and actual versions, are relevant.
    Aside from the meta data covered by our guidelines, we aim to include insights on discovered Linux kernels and \acp{ISA}.
    Filtering of such information allows researchers to create sub-corpora that suit their specific scenario.
    OSS samples are not included.
\end{itemize}

\noindent \circled{B} \textbf{Manufacturer Selection \& Device Classes.} We did not model market distributions in our corpus because there was no such information available.
Instead, we have selected ten manufacturers of consumer network appliances by subjectively perceived prevalence, sample availability, and portfolio:
The broader the portfolio of device classes, the better for heterogeneity.
The manufacturers are:
ASUS, AVM, D-Link, EDIMAX, EnGenius, Linksys, NETGEAR, TP-Link, TRENDnet, and Ubiquiti.
Examples for covered device classes are routers, switches, IP cameras, NAS systems, and network printers.

\noindent \circled{C} \textbf{Sample Acquisition.} For each manufacturer's page, we created scrapers with Scrapy~\cite{scrapy} v2.9.0.
We interfaced with Selenium and geckodriver v0.30.0 to render JavaScript pages.
For each manufacturer, we navigated through the steps of browsing the portfolio, inspecting devices, opening the support, and downloading the firmware by hand.
Along the way, we dissected page layouts to identify and implement relevant interactions for automation.
We pointed the scrapers to the manufacturer portfolio overviews as root.
Then, we let them traverse through the pages to collect all firmware.
If there were multiple hardware or firmware versions, we downloaded images for all of them.
If the vendors did not provide historical versions, we implemented fallback scrapers that use the WaybackMachine~\cite{wayback_cdx} to get samples from \texttt{archive.org}.

\noindent \circled{D} \textbf{Meta Data Acquisition.}
The scrapers extract as much meta data as possible from the manufacturers.
This includes the release date, version, manufacturer, model, and device class.
As for classes, we manually assigned labels to each product line.
In total, there are 22 labels: switch, router, ipcam, repeater, mesh, controller,
accesspoint, powerline, modem, power\_supply, wifi-usb,
recorder, nas, phone, board, kvm, converter, san,
printer, media, encoder, and gateway.
We saved download links and calculated file hashes like SHA256, as well as the fuzzy SSDeep and TLSH.

\noindent \circled{E} \textbf{Firmware Deduplication.}
We deduplicate firmware before unpacking by calculating the SHA256 sample hash.
In total, we collected 14,583 unique but packed firmware files.

\noindent \circled{F} \textbf{Replicable Unpacking.}
We found that the open source \ac{FACT}~\cite{fact} addresses many issues of replicable unpacking.
The tool gained attention in the non-academic security community and
provides plugin-based firmware processing for automated unpacking and static analyses.
Plugins register themselves to process files based on their MIME type.
Results are attached to each file as meta data.
The contents of a firmware are both subject to analysis and further recursive unpacking, which means that the tool selects the appropriate unpacker \textit{and protocols the unpacking results on file-basis}.
The core has unpackers for over 110 file types; some of them are reverse-engineered by the community, increasing the odds of unpacking success in comparison to, e.g., the academic's default binwalk~\cite{binwalk}, which is included as fallback when other approaches fail.
All collected images were uploaded to \ac{FACT} v4.2-dev.
We provide Vagrant~\cite{vagrant} recipes to easily deploy \ac{FACT} instances.
Our shared scripts ingest the shared \ac{LFwC} meta data to replicate corpus downloading and unpacking.
If the included links are dead, a fallback searches on the WaybackMachine.

\noindent \circled{G} \textbf{Content Deduplication.}
\ac{FACT} implements hash deduplication to find duplicate files across all firmware images.
Files can be correlated, similar to Costin et al.~\cite{costinLargeScaleAnalysisSecurity2014}.
Duplicates can be pruned to clean analysis results, which solves the issue on duplicate contents in firmware corpora (cf.~\cref{sec:literature:key_observations}).

\noindent \circled{H} \textbf{\acp{ISA} \& Kernels.}
We used \ac{FACT}'s architecture\_detection and software\_components plugins.
The former identifies \acp{ISA} in firmware samples.
It finds device trees, Linux kernel build configurations, and ELF file headers to skim them for common architecture identifiers such as arm64 or mips32el.
The latter uses YARA~\cite{yara} rules to find Linux banner version strings embedded in kernel builds.
This provides evidence that the firmware contains a Type-I or -II Linux system.
We export the results to complement the set of meta data included in~\ac{LFwC}.

\noindent \circled{I} \textbf{Verify Unpacking Success.}
After unpacking, we collected the files of all 14,583 firmware samples.
We marked a firmware image as successfully unpacked if we were able to find common Linux path components of extracted files.
Examples are \textit{/bin/}, \textit{/lib/}, and \textit{/var/}.
The full list is included in~\cref{fig:corpus_creation}.
This approach was chosen because the absence of, e.g., a Kernel banner from step \circled{H} does not imply failures:
Samples must not contain full root file systems, as it can also be an incremental device update.
We verified that 10,913 samples were successfully extracted.
\ac{LFwC} only contains these images and is, together with \ac{FACT}, a fully unpackable corpus.
The remaining 3,670 \textit{failed} images are a separate data set we share.

\noindent \circled{J} \textbf{Vulnerability Ground Truth.}
We tried to establish vulnerability ground truth.
The approach is twofold:
First, we map all devices in the corpus to RouterSploit~\cite{routersploit}, which consolidates exploits for network appliances.
There is meta data attached to each exploit, linking to possibly affected firmware versions and security advisories.
This yielded over 800 initial matches between samples and exploits.
As the meta data in RouterSploit is not complete, we manually inspected the matches and compared version strings with source security advisories.
We distilled 105 samples from the list with possibly applicable remote exploits, e.g., CVE-2017-5521 and CVE-2013-3093.
We note that version data in security advisories is often faulty~\cite{unreliable_cpe} and
effective matching is an open research topic~\cite{version_matching}.
To find additional evidence, we manually verified the presence of CVE-2016-10177 to CVE-2016-10186 in \ac{LFwC} using static analysis.
This set of vulnerabilities affected versions of the D-Link DWR-932B in 2016.

\subsection{Brief Insights on Corpus Composition}
\label{sec:corpus:composition}

\begin{table}[t]
    \caption{LFwC: Corpus Statistics Overview}
    \label{tab:corpus:stats}
    \centering
    \scalebox{0.860}{
    \begin{tabularx}{\columnwidth}{rccccc}
        \toprule
        \textbf{Manufact.}    & \textbf{Samples}                    & \textbf{Devices}         & $\mathbf{\frac{\text{\textbf{Samples}}}{\text{\textbf{Device}}}[\mu]}$            & $\frac{\text{\textbf{Size}}}{\text{\textbf{Sample}}}[\mu]$ & $\mathbf{\frac{\text{\textbf{Files}}}{\text{\textbf{Sample}}}[\mu]}$\\
        \midrule
        AVM                   & 797                        & 201                     & 3.97                              & 22 MiB                          & 2,194 \\
        TP-Link               & 1,163                      & 477                     & 2.44                              & 14 MiB                          & 1,446 \\
        ASUS                  & 1,647                      & 205                     & 8.03                              & 39 MiB                          & 3,780 \\
        D-Link                & 1,929                      & 458                     & 4.21                              & 19 MiB                          & 1,234 \\
        EDIMAX                & 200                        & 155                     & 1.29                              & 4  MiB                          & 395   \\
        EnGenius              & 143                        & 61                      & 2.34                              & 9  MiB                          & 1,144 \\
        Linksys               & 308                        & 166                     & 1.86                              & 13 MiB                          & 1,363 \\
        NETGEAR               & 2,580                      & 270                     & 9.56                              & 24 MiB                          & 2,145 \\
        TRENDnet              & 752                        & 191                     & 3.94                              & 9  MiB                          & 826   \\
        Ubiquiti              & 1,394                      & 181                     & 7.70                              & 78 MiB                          & 11,676\\
        \midrule
        \textbf{Total}        & 10,913                     & 2,365                   & 4.61                              & 29 MiB                          & 3,219 \\
        \bottomrule
    \end{tabularx}
    }
\end{table}

We give brief insights on \ac{LFwC}'s sample composition.
For more detail information, e.g., included \acp{ISA} and kernel distributions, we refer to Appendix~\ref{appendix:corpus}.

\cref{tab:corpus:stats} shows the file statistics across manufacturers.
\ac{LFwC} contains firmware images for 2,365 different devices and provides, on average, between four and five firmware versions per device.
The mean sample size is 29 MiB and the mean number of included files is 3,219.
Yet, the relatively large samples by Ubiquiti skew these numbers.
With 2,580 samples (24\%), NETGEAR is most represented.
D-Link~(18\%), ASUS~(15\%), Ubiquiti~(13\%), and TP-Link~(11\%) follow..

\begin{figure}[t!]
    \begin{center}
        \includegraphics[width=0.9\columnwidth]{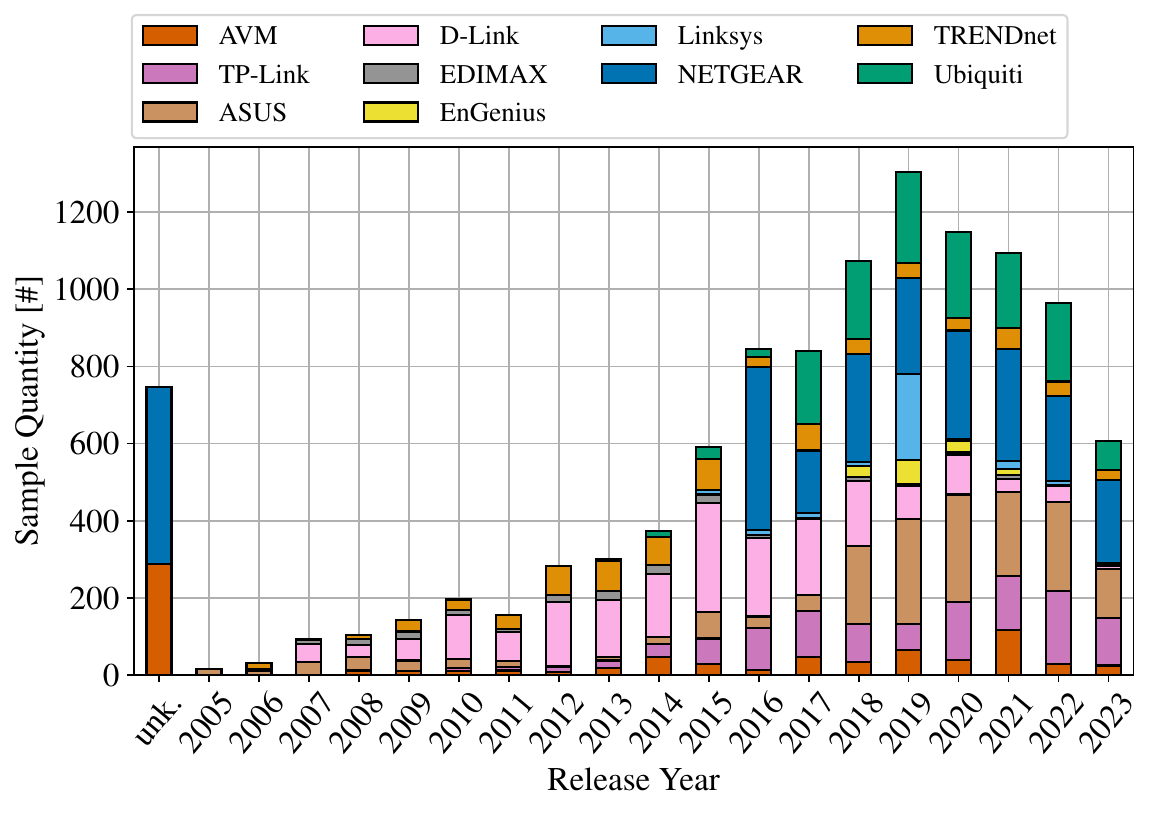}
        \caption{\ac{LFwC} firmware distribution per release date. For 747 samples, our scrapers could not extract any release date from the sources.}
        \label{fig:corpus:release_dates}
    \end{center}
\end{figure}

\cref{fig:corpus:release_dates} shows the samples per manufacturer across their year of release.
\ac{LFwC} has images from 2005 to June 2023, which proofs that it holds historical and recent samples.
From years 2005 to 2019, we observe exponential growth in sample quantities.
This makes sense considering the market growth over the past 20 years and changing patch behavior due to security awareness.
Starting with the COVID-19 pandemic, there is a decrease of samples between 2020 and 2022.
The data for 2023 is incomplete, as the corpus was created in June 2023.
For 747 samples, we could not extract release dates.

\begin{table*}[t!]
    \centering
    \caption{One Year After Corpus Creation: Results for the Case Study on the Independent Replicability of LFwC through Meta Data}
    \label{tab:replicability}
    \scalebox{0.825}{
    \begin{tabularx}{\linewidth}{rcccccccccccc}
        \toprule
                              & \textbf{Samples} & \textbf{Samples}     & \multicolumn{2}{c}{\textbf{1: Link}} & \multicolumn{2}{c}{\textbf{2: Archive}} & \multicolumn{2}{c}{\textbf{3: VirusTotal}} & \multicolumn{2}{c}{\textbf{4: Manual}} & \multicolumn{2}{c}{\textbf{Missing}}\\
        \textbf{Manufact.}    & \textbf{LFwC}    & \textbf{Replicated}  & \textbf{Samples}  & \textbf{Ratio}   & \textbf{Samples}  & \textbf{Ratio}      & \textbf{Samples}  & \textbf{Ratio}         & \textbf{Samples}  & \textbf{Ratio}     & \textbf{Samples}  & \textbf{Ratio}\\
        \midrule
        AVM                   & 797              & 790                  & \textbf{731}      & \textbf{0.91}    & 13                & 0.02                & 34                & 0.04                   & 12                & 0.02               & 7                 & 0.02\\
        TP-Link               & 1,163            & 1,163                & \textbf{1,163}    & \textbf{1.00}    & 0                 & 0.00                & 0                 & 0.00                   & 0                 & 0.00               & 0                 & 0.00\\
        ASUS                  & 1,647            & 1,642                & \textbf{1,633}    & \textbf{0.99}    & 7                 & $<$0.01             & 2                 & 0.01                   & 0                 & 0.00               & 5                 & 0.00\\
        D-Link                & 1,929            & 1,927                & \textbf{1,911}    & \textbf{0.99}    & 0                 & 0.00                & 2                 & 0.01                   & 14                & 0.01               & 2                 & $<$0.01\\
        EDIMAX                & 200              & 200                  & \textbf{200}      & \textbf{1.00}    & 0                 & 0.00                & 0                 & 0.00                   & 0                 & 0.00               & 0                 & 0.00\\
        EnGenius              & 143              & 143                  & \textbf{143}      & \textbf{1.00}    & 0                 & 0.00                & 0                 & 0.00                   & 0                 & 0.00               & 0                 & 0.00\\
        Linksys               & 308              & 308                  & \textbf{308}      & \textbf{1.00}    & 0                 & 0.00                & 0                 & 0.00                   & 0                 & 0.00               & 0                 & 0.00\\
        NETGEAR               & 2,580            & 2,564                & \textbf{2,551}    & \textbf{0.98}    & 0                 & 0.00                & 12                & $<$0.01              & 1                 & $<$0.01            & 16                & $<$0.01\\
        TRENDnet              & 752              & 752                  & \textbf{752}      & \textbf{1.00}    & 0                 & 0.00                & 0                 & 0.00                   & 0                 & 0.00               & 0                 & 0.00\\
        Ubiquiti              & 1,394            & 1,394                & \textbf{1,394}    & \textbf{1.00}    & 0                 & 0.00                & 0                 & 0.00                   & 0                 & 0.00               & 0                 & 0.00\\
        \midrule                                                                                                                                                                                                                                                         
        \textbf{Total}        & 10,913           & 10,883               & \textbf{10,786}            & \textbf{0.99}    & 20       & $<$0.01                & 50                & $<$0.01                   & 27                & $<$0.01               & 30                & $<$0.01\\
        \bottomrule
    \end{tabularx}
    }
\end{table*}

\subsection{Case Study I: Proof of Replicability}
\label{sec:corpus:replication}
We share meta data for replication.
Thus, there is an inherent risk that \ac{LFwC}'s included download links become inaccessible over time.
The requirements in~\cref{sec:reqs} propose to include file- and device-related data, e.g., product data and hashes.
This information helps to acquire files from other sources.

We conducted a case study to investigate the replicability of \ac{LFwC}:
In June 2024, one year after creating the corpus, we simulated the scenario of an independent researcher that only uses the shared meta data and tools to reconstruct \ac{LFwC}.
We divided the process into four phases:

\begin{enumerate}
    \item Pass the shared meta data to our download tool, which obtains the samples using the original links.
    \item Then, use the archive.org fallback (cf.~\cref{sec:corpus:creation}).
    \item Search firmware images on VirusTotal~\cite{virustotal} via hashes.
    \item Use the device name, file name, and version data to search for samples on forums and archives.
\end{enumerate}

The first three steps were automated and the fourth was manual.
We noted the phase of successful acquisition for each sample and collected the results.
The total size of packed samples is 353~GiB.
We used a 1 Gbps Internet connection.

\cref{tab:replicability} shows the results per manufacturer and phase:
After one year, we were able to successfully replicate 99.73\% of \ac{LFwC} (10,883 out of 10,913 samples) within five hours.
10,786 samples were acquired from the included direct download links, which left us with the task to recover 177 samples from alternative sources.
Entering the fallbacks, we found 13 missing AVM and seven missing ASUS images on archive.org.
Another 50 were found on VirusTotal~\cite{virustotal}, most of them from AVM (34) and NETGEAR (12).
Then, we invested roughly 40 minutes for manual search on the Internet and gathered another 27 samples, of which we were able to verify hash identity.
Most links in this category are dead because the manufacturers updated their devices and moved the older firmware packages to another URL, which was recovered.
30 samples could not be found within the 40 minutes time span of manual acquisition.

Overall, the results provide evidence that independent researchers can efficiently replicate \ac{LFwC} through its meta data and automated tools; with reasonable manual effort.
The fact that 99\% of the samples could still be acquired through their originally scraped links, is a positive result.
However, one should expect that this rate decreases over time and more samples must be collected through fallback methods.

\subsection{Case Study II: Proof of Scientific Corpus Utility}

This brief case study demonstrates the scientific utility of \ac{LFwC}:
Together with \ac{FACT}, we successfully perform a large-scale trend analysis using static methods.

Similar to Yu et al.~\cite{yuBuildingEmbeddedSystems}, we raise the following question:
\textit{How did the use of binary hardening methods for ELF files change over time in the firmware corpus at hand?}
We focus on five compiler-based user space methods:
Canaries, Non-Executable Stacks (NX), Relocation Read-Only (RELRO), Position-Independent Code (PIC), and Fortify Source.
Yu et al.~\cite{yuBuildingEmbeddedSystems} explain these hardening methods in detail.

\begin{table}[t!]
    \centering
    \caption{ELF Binaries Extracted from LFwC}
    \label{tab:elf_files}
    \scalebox{0.83}{
    \begin{tabular}{rcccccccc}
        \toprule
        & \multicolumn{4}{c}{\textbf{Not Deduplicated [\#k]}} & \multicolumn{4}{c}{\textbf{Deduplicated [\#k]}}\\
        \textbf{Arch}     & \textbf{Execs}   & \textbf{Libs}    & \textbf{Objs}  & $\mathbf{\Sigma}$ & \textbf{Execs} & \textbf{Libs}  & \textbf{Objs}   & $\mathbf{\Sigma}$\\ \midrule
        \textbf{ARM}      & 638.1   & 630.8   & 30.0  & 1,568.9 & 102.6  & 56.4  & 34.5 & 193.4 \\
        \textbf{MIPS}     & 416.7   & 397.5   & 90.6  & 904.8   & 87.6   & 43.3  & 26.6 & 157.5 \\
        \textbf{x86}      & 26.8    & 34.4    & 6.0   & 67.3    & 3.8    & 4.8   & 1.6  & 10.2  \\
        \textbf{Other}    & 13.2    & 6.5     & 11.1  & 30.8    & 4.3    & 1.0   & 1.7  & 7.0   \\
        \midrule
        $\mathbf{\Sigma}$ & 1,094.8 & 1,069.3 & 407.8 & \textbf{2,571.8} & 198.3  & 105.5 & 64.4 & \underline{\textbf{368.2}}\\
        \bottomrule
    \end{tabular}
    }
    \textbf{Execs:} x-pie-executable, x-executable. \textbf{Libs:} x-sharedlib, x-archive.\\\textbf{Objs:} x-object.
\end{table}
\begin{figure}[t!]
    \begin{center}
        \includegraphics[width=0.85\columnwidth]{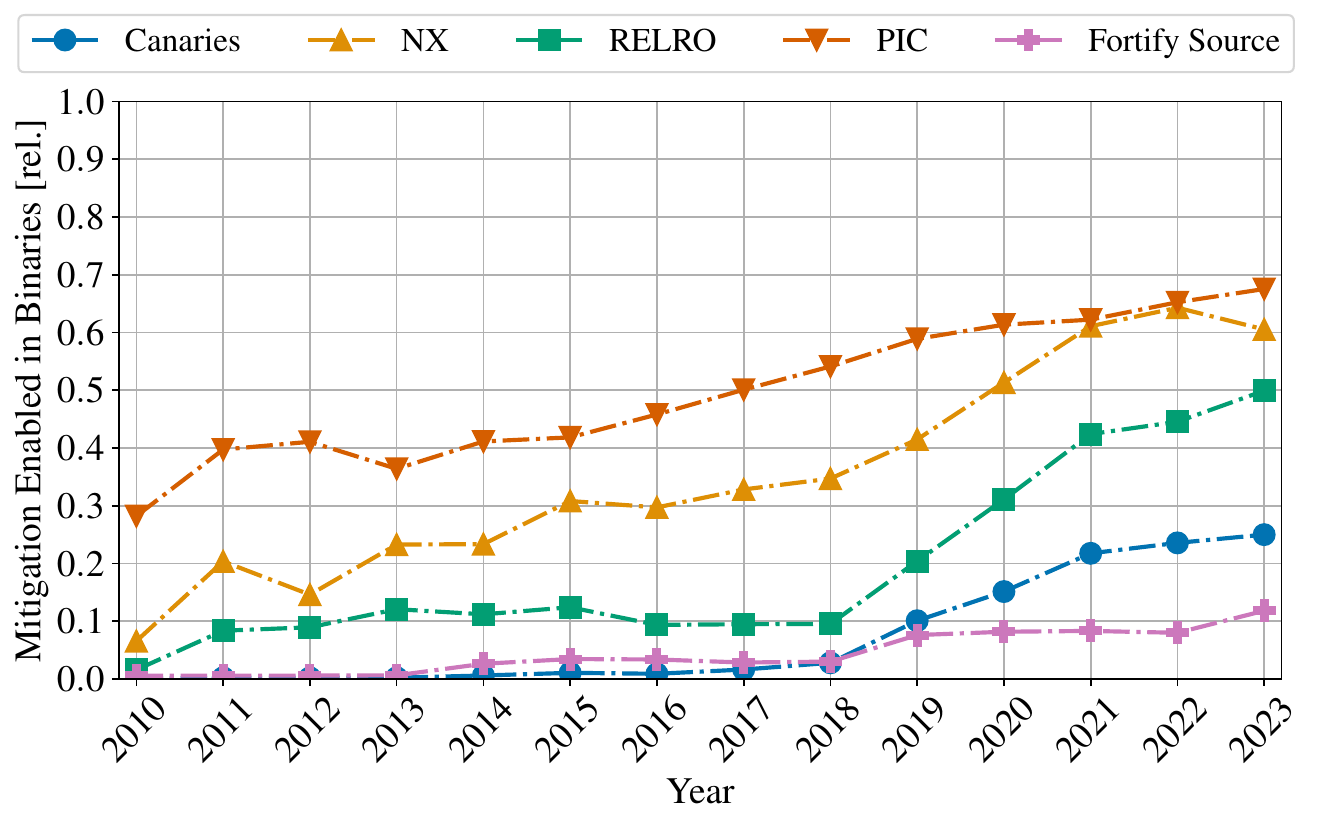}
        \caption{ELFs with enabled hardening techniques per firmware release year.}
        \label{fig:corpus:mitigations}
    \end{center}
\end{figure}

With \ac{LFwC} fully unpacked in \ac{FACT} v4.2-dev, we query the tool's API to aggregate all included files that have common ELF MIME types.
We purged all kernel objects (.ko extension) and images (\ac{FACT} Linux kernel detection) from the data set.

\cref{tab:elf_files} shows the data and its MIME types:
We extracted 2.6m ELFs from all 10,913 images.
The majority uses ARM or MIPS.
\ac{FACT}'s deduplication helped to create a final data set of 368.2k unique ELFs across all firmware images.

We applied checksec v2.6.0~\cite{checksec} to the 386.2k files to detect the presence of binary hardening methods in ELF headers.
We traced back each ELF to its origin firmware, associate the firmware's release date, and grouped the analysis results by year.
For each method, \cref{fig:corpus:mitigations} plots the fraction of ELFs included in firmware from 2010 to 2023 where the method is enabled.
From 2010 to 2023, usage grows steadily for all hardening methods.
While PIC is most often enabled and grew almost linearly from 2010 (30\%) to 2023 (67\%), we observe an accelerating trend for all other methods starting with 2018.
As of 2023, NX (60\%) almost catched up to PIC, followed by RELRO (50\%), Canaries (25\%), and Fortify Source (12\%).
Overall, we can observe a positive trend in our data set.
This is reasonable, as security awareness grew and resources needed to use these methods became more affordable.

This contradicts the results by Yu et al.~\cite{yuBuildingEmbeddedSystems} from 2022:
As shown in Figure 4 of their paper, they observe almost 100\% NX usage, while Canaries fluctuate, and the other methods have almost no change in adoption rates.
The authors note that their data set does not seem balanced, as five out of 34 manufacturers provide 78\% of the ELF binaries.
As Synology makes up for 46\% of all ELFs, changes in their build chain 
can significantly impact results.

\begin{figure}[t!]
    \begin{center}
        \includegraphics[width=0.85\columnwidth]{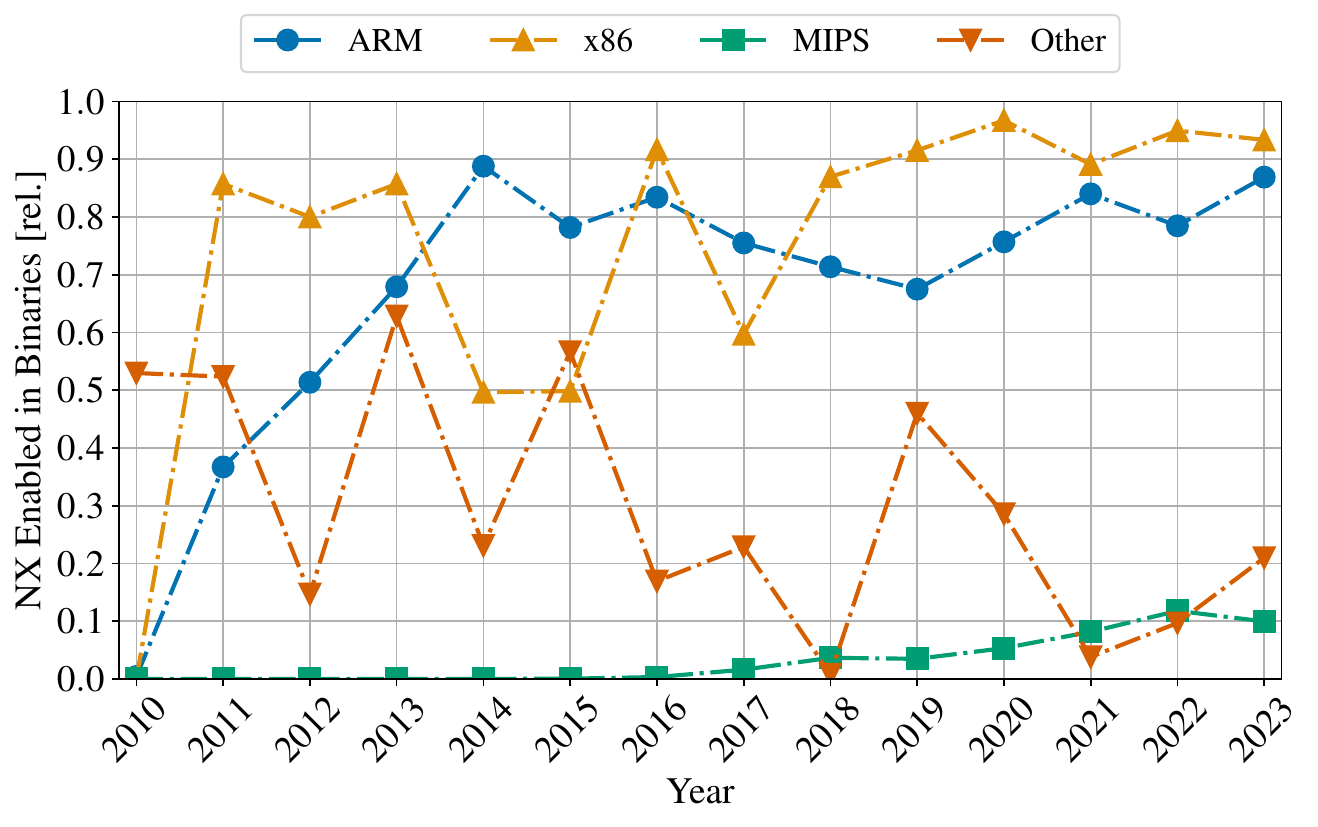}
        \caption{ELFs with enabled NX bit per release year and \ac{ISA}.}
        \label{fig:corpus:nx_by_arch}
    \end{center}
\end{figure}

Also, we can use \ac{LFwC} to show that the 100\% NX adoption rate from Yu et al.~\cite{yuBuildingEmbeddedSystems} can not be representative.
NX requires CPU support, as the hardware has to ultimately protect memory regions from execution.
ARM and x86 families support NX since decades, but MIPS specifications only introduced this feature recently as eXecute Inhibit (XI)~\cite{hrsr}.
As there is a certain time-to-market for each specification, we can assume that fewer MIPS devices support this feature than x86 and ARM devices.
\cref{fig:corpus:nx_by_arch} shows the adoption rate of NX in our corpus by architecture over the past decade.
As expected, the ARM and x86-based ELFs almost all adopted NX as of 2023.
However, for MIPS, NX adoption only started in 2017.
Six years later, the adoption rate is at 10\%, which is reasonable.
As we show in~\cref{tab:elf_files}, 42\% of the ELF files in our corpus are MIPS and \acp{ISA} other than ARM and x86 are negligible.
At the same time, 41\% of the ELF files are MIPS in the corpus by Yu et al.~\cite{yuBuildingEmbeddedSystems}.
Thus, both data sets should be comparable, but the results vastly deviate.
We invite interested researchers to further explore \ac{LFwC} and investigate the observations we made as part of this short case study, in which we showed that our corpus can be used for static analyses.

\subsection{Ethical Discussion on Construction and Distribution}
\label{sec:corpus:ethical_discussion}
For ethical firmware collection, we taught our scrapers to obey to the \texttt{robots.txt} of manufacturers.
We assume that said file correctly implements any restrictions mentioned in their ToS/EULA.
We also reduced infrastructure load by throttling requests and execution threads.
The replication scripts we provide are equally throttled.
We did not greedily mirror and process all information available on the manufacturer's pages.
These pages can include data, e.g., forum posts, that might fall under special data protection laws.
Thus, we designed our scrapers in a way that steers them directly towards the desired sample downloads over the product pages.
These were, alongside the specifically included meta data fields, the only data we persisted.
Furthermore, we argue that it is unethical to include detail FACT~\cite{fact} reports in the meta data we share.
These include security-relevant data, e.g., file system trees, hashes of included files, or complete lists of detected software components. This data provides a low entry barrier for malicious actors to perform large-scale presence checks of emerging zero day vulnerabilities across thousands of devices.
This brought us to the decision that it is responsible to control access to the meta data.
\ac{LFwC} is a scientific corpus that shall help researchers to evaluate methods to improve, and not jeopardize, the security of devices.

\subsection{Corpus Limitations \& Future Expansion}

Our creation methods imply various limitations.
First, we criticize the sample distribution:
As we only scrape easily accessible, network-centric samples, \ac{LFwC} does not include devices where acquisition is less scalable, more complex, and invasive.
Cisco IOS images, e.g., are gated by login pages.
We further limited the sources to ten manufacturers.
Thus, market distributions are not accurately modelled and trend analyses on samples might lack fidelity.
OSS samples are also not included.
Generally, \ac{LFwC} is limited to Type-I and -II Linux images.
Their distribution is unknown:
Device-specifics like timing constraints might incentivize manufacturers to implement user space applications in kernel space, which blurs the lines between the two classes (cf. Appendix~\ref{appendix:types}).
A detail content analysis would be required for all 10,913 samples to distinguish, which is infeasible within the scope of this paper.

Second, \ac{LFwC}'s compatibility is limited to static methods.
We can not test the presence of execution parameters that are possibly needed without having concrete dynamic scenarios at hand.
This does not mean that it is unsuitable for dynamic analyses, but that we did not verify \ac{LFwC}'s compatibility.

Third, we note that \ac{FACT} uses static heuristics for unpacking, verification, and meta data acquisition.
We also mapped Routersploit ground truth statically and manually.
Thus, there is a chance of false-positive findings in the corpus meta data.

Finally, we observe that all shared meta data for replication remains volatile and the relevancy of samples fades as the corpus ages.
Thus, we plan to provide at least annual updates that refresh the links and add new manufacturers or devices to the corpus.
As our scrapers are open source, we invite researchers to join us in this task.

\section{Conclusion}

We brought attention to an important, but often overlooked aspect of binary vulnerability research: The creation of representative and replicable firmware corpora for sound and independently verifiable experiments.
For this purpose, we pinpointed eight challenges that can significantly affect corpus creation.
We used them to derive a strict framework of corpus requirements, nurtured by 16 measures.
The goal was to give broadly applicable and practical guidelines that can be used to improve soundness -- given current copyright laws and unpacking barriers.
We revised the status quo of corpus creation practices through a systematic literature review on top tier research.
We showed that our framework serves its purpose, as it helps to find and avoid many methodical step stones:
Missing meta data, incomplete documentation, and inflated corpus sizes blur visions on representativeness and hinder replicability.
Our measures show practical ways, like an extended set of meta data, that one could implement to improve corpus soundness.

The requirements led to the creation of~\ac{LFwC}, a new Linux firmware corpus with rich meta data, high-quality deduplicated samples, and verified contents.
It is a valuable addition to the community, as it demonstrates verified replicability, shows its scientific utility, and takes the needed space to transparently document creation and composition.
Together with our requirements, it is a step towards scientifically sound corpora.
We plan yearly updates to preserve relevance and replicability.

\bibliographystyle{IEEEtranS}
\bibliography{bibliography}

\appendices

\section{Firmware Taxonomy}
\label{appendix:types}

Embedded devices are diverse:
They operate in many environments, are highly specialized, and are bound to application-specific constraints or cost functions~\cite{challenges_firmware_rehosting}.
This leads to many combinations of hardware layouts, \acp{ISA}, software stacks, and peripherals.
Thus, heterogeneity shapes the firmware corpora and analysis methods in research.

To better describe firmware and its heterogeneous property space in this paper, we adopt the taxonomy by Muench et al.~\cite{muenchWhatYouCorrupt2018}.
They define four types on the axes of \ac{OS} abstraction and specialization.
This appendix describes them in detail:

\begin{itemize}[align=left, wide=0em, leftmargin=1em]
    \item[\textbf{Type-0}] \emph{Conventional Desktop and Server Systems.}
    This type describes common desktop or~server \acp{OS}, e.g., Windows, Linux, and Mac OS.
    They ship rich user environments and general purpose kernels that add many hardware abstractions and features like scheduling or file systems.
    Type-0 has many functions and high modularity, but requires certain components and more resources than other types.
    It is not strictly firmware~\cite{muenchWhatYouCorrupt2018}, but influences the other types.
    Dominant \acp{ISA} are x86 and aspiring ARM platforms.
    \item[\textbf{Type-I}] \emph{General Purpose Embedded Systems}.
    Type-I retrofits Type-0 for embedded use:
    Because manufacturers know device purpose and hardware, they can strip unused components and add proprietary code.
    This allows less powerful devices to host Type-0 applications.
    Routers, e.g., may use networking from Type-0 \acp{OS}, but do not need media codecs.
    Linux is a common \ac{OS} in Type-I systems~\cite{costinLargeScaleAnalysisSecurity2014,chenAutomatedDynamicAnalysis2016}.
    Here, manufacturers couple stripped kernels with slim user space environments like BusyBox~\cite{challenges_firmware_rehosting}.
    Other popular Type-I devices use Windows IoT, Android, or iOS.
    MIPS and ARM are common \acp{ISA}~\cite{costinLargeScaleAnalysisSecurity2014}.
    \item[\textbf{Type-II}] \emph{Special Purpose Embedded Systems}.
    Type-II runs on single purpose systems common in, e.g., industrial, automotive, or healthcare settings. 
    Respective devices are robotic arms, wallboxes, or ECGs.
    Type-II handles requirements where Type-I fails, e.g.:
    There is no \acl{MMU} or there is limited processing power.
    Hardware layouts or special components may be unsupported by Type-I.
    There could be no file system, or tasks may be too time-critical for general purpose scheduling.
    Kernel and user space still exist, but lines become blurry~\cite{muenchWhatYouCorrupt2018}.
    MIPS and ARM are present, but there are plenty alternatives.
    Example Type-II \acp{OS} are VxWorks, Zephyr OS, $\mu$Clinux, and FreeRTOS.
    \item[\textbf{Type-III}] \emph{Bare-metal Embedded Systems}.
    Type-III, or bare-metal, systems have few to no \ac{OS} abstraction:
    The firmware is a monolithic executable that shows no separation between kernel and user space.
    Applications execute in a single loop and have direct access to hardware~\cite{muenchWhatYouCorrupt2018}.
    Being minimalistic and low level, Type-III shows highest flexibility and least requirements.
    Thus, it runs on controllers or other integrated circuits.
    It is also used in highly constrained \ac{IoT} devices, sensors, and actuators.
    The code can be custom or proprietary, but may use \ac{OS}-like libraries for protocol stacks, interrupt handlers, and memory management.
    Example libraries are Contiki-NG and Mbed OS.
\end{itemize}

\section{Catalog of Criteria for the Literature Review}
\label{appendix:criteria}
This appendix provides a catalog of criteria that we applied to each paper to derive the literature review results in~\cref{fig:literature:req_results}.

\subsection{General Information}
\begin{itemize}
    \item If papers reused existing corpora, we considered all information on corpus creation practices from the sources and their artifacts as well.
    \item We only considered data sets with real-world samples.
    \item When papers used multiple real-world firmware corpora, we considered all of them.
    \item If a measure did not fit a paper's research question, we explicitly marked the data point as not applicable~(\notapplicable). Example: An \ac{HIL}-based method like IoTFuzzer~\cite{iotfuzzer} does not need sample unpacking.
    Thus, neither unpacked sample quantities, nor unpacking procedures can be documented.
\end{itemize}

\subsection{Fulfillment Criteria for all 16 Measures}

\noindent$\blacktriangleright$ \textbf{Packed \& Unpacked Samples, Manufacturers, Models, Device Classes, and ISAs.}
Let \textbf{\textit{M}} be one of the measures above. \textbf{\textit{M}} is ...
\begin{itemize}[align=left, wide=0em, leftmargin=1em]
    \item[\textbf{([0-9]+ / \yes)}] \textbf{fulfilled}, if the paper states concrete quantities of \textbf{\textit{M}} for \textit{all} samples ...
    \begin{enumerate}
        \item ... in any kind of text (main, appendices, footnotes)
        \item ... [OR] in figures or tables,
        \item ... [OR] in shared meta data/linked repositories.
        \item ... [OR] through complete sample lists in 1-3 that can be used to independently calculate the quantities.
    \end{enumerate}
    \item[\textbf{(\unclear)}] \textbf{partially fulfilled}, if the paper desribes \textbf{\textit{M}} ...
        \begin{enumerate}
        \item ... \textit{only for a subset} of samples as stated in \yes,
        \item ... [OR] \textit{imprecisely}, e.g., some round values~\cite{firmup}.
    \end{enumerate}
    \item[\textbf{(\no)}] \textbf{unfulfilled} in all other cases.
\end{itemize}

\noindent$\blacktriangleright$ \textbf{Deduplication, Unpacking, and Acquisition.}
Let \textbf{\textit{M}} be one of the measures above. \textbf{\textit{M}} is ...
\begin{itemize}[align=left, wide=0em, leftmargin=1em]
    \item[\textbf{(\yes)}] \textbf{fulfilled}, if the paper explicitly documents \textbf{\textit{M}} ...
    \begin{enumerate}
        \item ... in any kind of text (main, appendices, footnotes)
        \item ... [OR] in figures or tables,
        \item ... [OR] in shared meta data/linked repositories.
        \item ... [OR] through a complete list of sample data or scripts to independently verify or replicate \textbf{\textit{M}}.
    \end{enumerate}
    \item[\textbf{(\unclear)}] \textbf{partially fulfilled}, if the paper documents \textbf{\textit{M}} ...
        \begin{enumerate}
        \item ... \textit{only for a subset} of samples as stated in \yes,
        \item ... [OR] \textit{incomplete}, such that independent research must assume substeps for verification or replication.
    \end{enumerate}
    \item[\textbf{(\no)}] \textbf{unfulfilled} in all other cases.
\end{itemize}

\begin{figure*}[t]
    \begin{center}
        \includegraphics[width=0.85\textwidth]{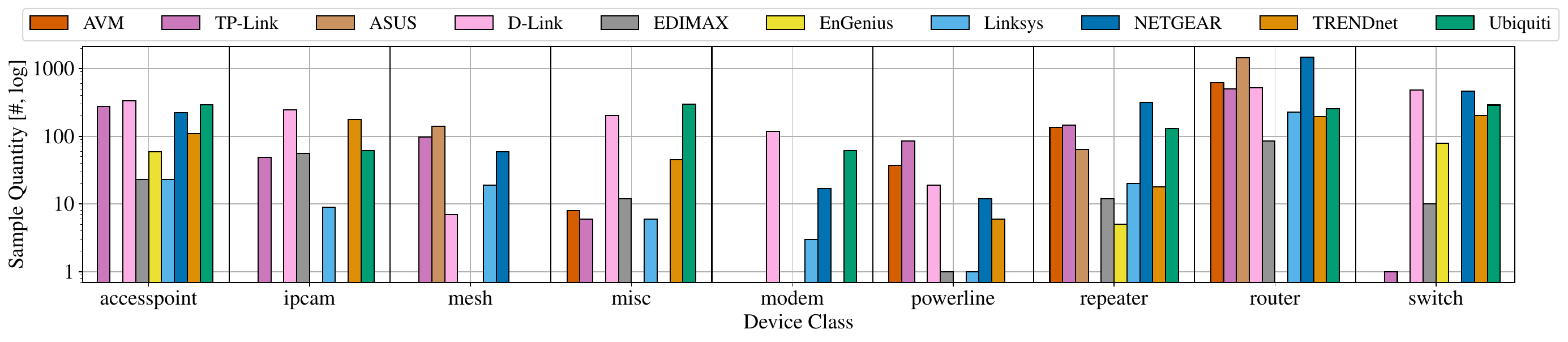}
        \caption{Distribution of device classes in~\ac{LFwC}. The three most prevalent classes are routers (49\%), switches (14\%), and access points (12\%). We bundled device classes with less than 150 samples into the meta class \textit{misc}. It contains: controller, board, converter, encoder, gateway, kvm, media, nas, phone, power\_supply, printer, recorder, san, and wifi-usb.}
        \label{fig:corpus:classes}
    \end{center}
\end{figure*}
\begin{figure*}[t]
    \begin{center}
        \includegraphics[width=0.85\textwidth]{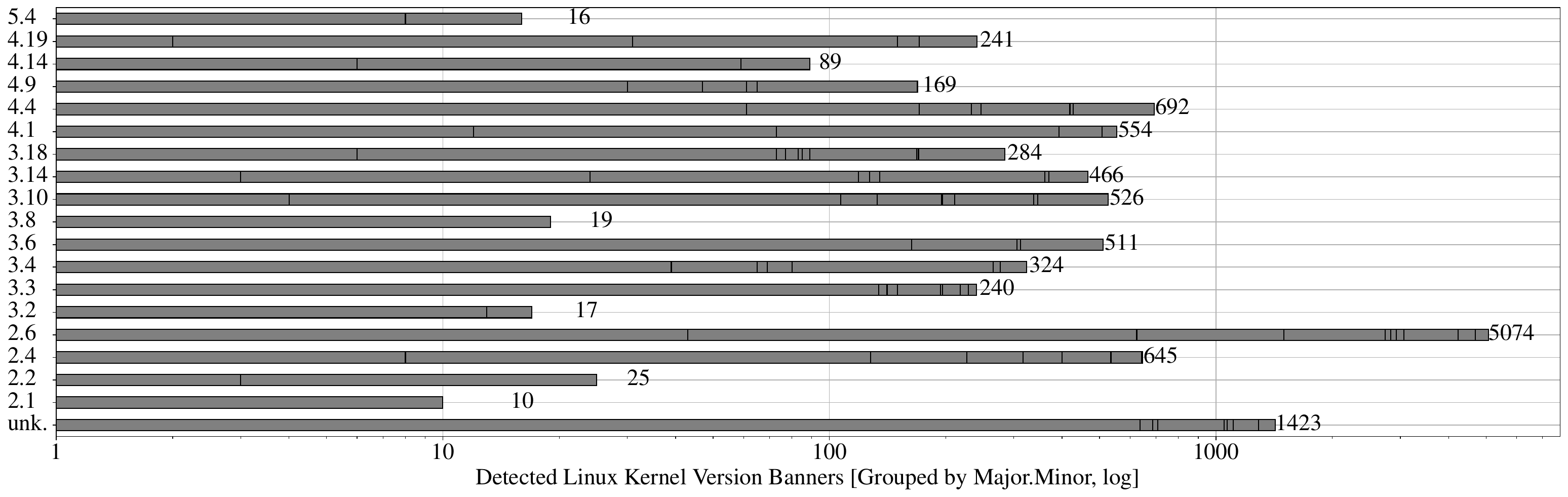}
        \caption{Detected Linux kernel banners in \ac{LFwC} samples. Versions range from 2.1 to 5.4, with Linux kernel 2.6 remaining the most prevalent version in our data set (51\%). In total, we found 9,901 kernel banners across all included images. We could not find a banner in 1,423 samples. Note that a sample can ship with multiple kernels hosted by subsystems of independent device components.}
        \label{fig:corpus:kernels}
    \end{center}
\end{figure*}
\begin{figure*}[t]
    \begin{center}
        \includegraphics[width=0.85\textwidth]{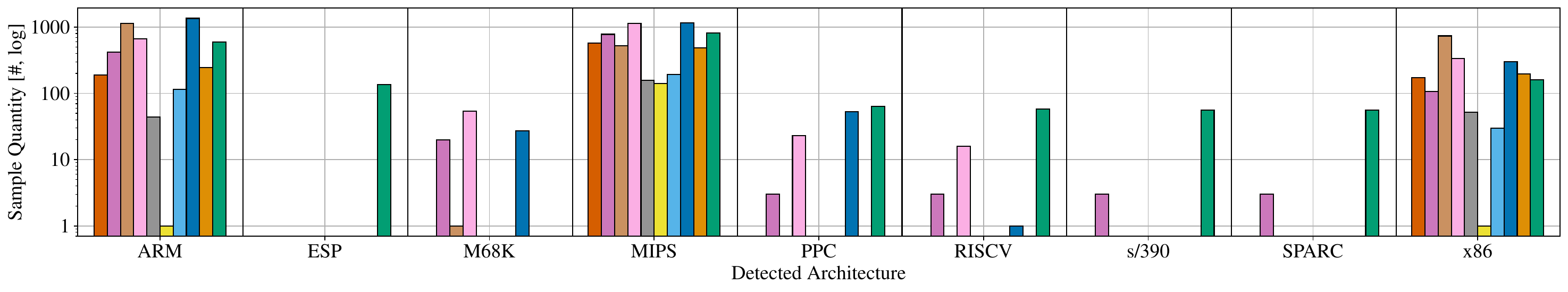}
        \caption{Distribution of the nine detected \acp{ISA} in \ac{LFwC} across all vendors on a logarithmic scale. The three most prevalent \ac{ISA} families are MIPS (5,993 samples), ARM (4,764), and x86 (2,095).
        There are 13,429 unique findings on \acp{ISA} across all samples, because included subsystems must not run the same \ac{ISA} as the main system.}
        \label{fig:corpus:architectures}
    \end{center}
\end{figure*}

\noindent$\blacktriangleright$ \textbf{Rel. Dates, Versions, Links, Hashes, and FW Types.}
Let \textbf{\textit{M}} be one of the file properties above. \textbf{\textit{M}} is ...
\begin{itemize}[align=left, wide=0em, leftmargin=1em]
    \item[\textbf{(\yes)}] \textbf{fulfilled}, if the paper provides \textbf{\textit{M}} for \textit{all} samples ...
    \begin{enumerate}
        \item ... in any kind of text (main, appendices, footnotes)
        \item ... [OR] in references,
        \item ... [OR] in figures or tables,
        \item ... [OR] in shared meta data/linked repositories.
        \item ... [OR] (Links and Hashes) through shared samples.
    \end{enumerate}
    \item[\textbf{(\unclear)}] \textbf{partially fulfilled}, if the paper fulfills \textbf{\textit{M}} \textit{only for a subset} of samples as stated in \yes.
    \item[\textbf{(\no)}] \textbf{unfulfilled} in all other cases.
\end{itemize}

\noindent$\blacktriangleright$ \textbf{Sample Selection Reasoning and Ground Truth}

\noindent Let \textbf{\textit{M}} be one of the above. \textbf{\textit{M}} is ...
\begin{itemize}[align=left, wide=0em, leftmargin=1em]
    \item[\textbf{\textit{(\yes)}}] \textbf{fulfilled}, if researchers provide reasons for sample selection or consciously include bug ground truth before evaluation. Reasons could be arbitrary and must not seem reasonable to everyone.
    However, they are explicitly discussed and, thus, documented and disputable. Examples: The method to evaluate only supports certain \acp{ISA}, manufacturers, or \acp{OS}.
    \item[\textbf{(\unclear)}] \textbf{partially fulfilled}, if there is no explicit selection reasoning, but possible selection motives are derivable from the paper. E.g., a paper rediscovers already known bugs in samples. A lack of documentation leaves the reader guessing if the researchers deliberately included samples with ground truth or not.
    \item[\textbf{(\no)}] \textbf{unfulfilled} in all other cases.
\end{itemize}

\section{LFwC: Insights on Corpus Meta Data}
\label{appendix:corpus}

\cref{fig:corpus:classes} shows the device classes per manufacturer.
With 5,328 (49\%) samples, routers are the most prevalent class in \ac{LFwC}, followed by switches (1,525/14\%) and access points (1,345/12\%).
The remainder of classes has a prevalence between 162 (powerline) and 851 (repeater) samples.
For illustration, we bundled the other 14 classes, with samples from 3 to 116, into \textit{misc} (569, or 5\%).

\cref{fig:corpus:kernels} shows the detected Linux kernel banners,
grouped by major and minor version.
We removed manufacturer colors as the scale introduced false visual cues of over- and underrepresented shares.
Versions range from 2.1 to 5.4, with 2.6 being most prevalent (5,074, or 51\%).
With 692 findings (7\%), version 4.4 is the second most prevalent, followed by 2.4 with 645 findings.
With 16 matches for version 5.4 and 10 matches for version 2.1, both extremes are rare in \ac{LFwC}.
In total, we found 9,901 banners in~\ac{LFwC}.
The above numbers are already sanitized, as \ac{FACT} found roughly 400 false positives which originated from a pptp-linux client binary with similar banner format.
Also, in some samples, we found multiple banners.
This is because images can contain multiple subsystems and, thus, kernels.
Cable modems embedded in routers, e.g., can run independent kernels interfacing with the main system.
This data is present because the updates for sub components are managed by the main system.
In general, we found that the existence of these subsystems in firmware images is often overlooked in related work on large scale analyses.

Finally,~\cref{fig:corpus:architectures} shows the nine detected \acp{ISA}.
There are 13,429 unique \ac{ISA} findings hinting at the subsystems embedded into devices.
In 5,993 images, we found MIPS, making it the most common \ac{ISA} in \ac{LFwC}.
ARM was found in 4,764 images.
The third group is x86 with findings in 2,095 images -- most of them in powerful devices like managed switches and enterprise routers.
With findings in 59 to 143 samples, the remaining \acp{ISA}, usually found in micro-controller sub-components, are underrepresented.

\end{document}